\documentclass[
amsmath,amssymb,
aps, prl,
twocolumn
]{revtex4-2}

\usepackage{graphicx}
\usepackage{dcolumn}
\usepackage{bm}
\usepackage[dvipsnames]{xcolor}
\usepackage{epstopdf}
\usepackage[labelfont=bf]{subcaption}

\usepackage{tikz}
\usetikzlibrary{decorations.pathreplacing}
\usepackage{subcaption}

\def\beq{\begin{eqnarray}}
\def\eeq{\end{eqnarray}}
\def\be{\begin{equation}}
\def\bel{\begin{equation}\label}
\def\beel{\begin{eqnarray}\label}
\def\ee{\end{equation}}
\def\eq{&=&}
\def\ct{\cite}
\def\l{\left}
\def\r{\right}
\def\bm{\begin{math}}
\def\me{\end{math}}
\def\bi{\bibitem}
\def\vr{\vec r}
\def\om{\omega}
\def\lt{L(t)}
\def\al{a(L,T)}
\def\hf{\frac{1}{2}}

\def\la{\langle}
\def\ra{\rangle}
\def\lbr{\left [}
\def\rbr{\right ]}
\def\del{\partial}

\def\ul{\underline}
\def\etal{{\it et al.}}
\def\lra{\leftrightarrow}
\def\rar{\rightarrow}
\def\lb{\label}
\def\q{\quad}
\def\qq{\qquad}
\def\d{\delta}
\def\D{\Delta}

\newcommand \nn {\nonumber}
\newcommand \bei {\begin{itemize}}
\newcommand \eei  {\end{itemize}}

\renewcommand{\leq}{\leqslant}
\renewcommand{\geq}{\geqslant}

\begin{document}
\bibliographystyle{apsrev}

\title{Hydrodynamic equations for space-inhomogeneous aggregating fluids with first-principle
kinetic coefficients
}

\author{A. I. Osinsky$^{1,2}$}
\author{N. V. Brilliantov$^{1,3}$}
\affiliation{$^{1}$ Skolkovo Institute of Science and Technology, 121205, Moscow, Russia}
\affiliation{$^{2}$ Marchuk Institute of Numerical Mathematics of Russian Academy of Sciences, Moscow, 119333 Russia}
\affiliation{$^{3}$ Department of Mathematics, University of Leicester, Leicester LE1 7RH, United Kingdom}

\begin{abstract}
We derive from the first principles
new hydrodynamic equations --
Smoluchowski-Euler equations for  aggregation kinetics in space-inhomogeneous fluids with fluxes. Starting from Boltzmann equations, we obtain microscopic expressions for aggregation rates for clusters of different sizes and observe that they significantly differ from currently used phenomenological rates. Moreover, we show that for a complete description of aggregating systems, novel kinetic coefficients are needed. They share properties of transport and reaction-rate coefficients;  for them we report microscopic expressions.
For two representative examples -- aggregation of particles at sedimentation and aggregation after an explosion we numerically solve Smoluchowski-Euler equations and perform Direct Simulation Monte Carlo (DSMC).  We find that while the new theory agrees well with DSMC results, a noticeable difference is observed for the phenomenological theory. This manifests the unreliability of the currently used phenomenological theory and the need to apply new, first-principle equations.
\end{abstract}
\maketitle
{\em Introduction.} Aggregation is ubiquitous in natural systems and widely used in technological processes \cite{Smo16,muller1928allgemeinen,Smo17,Chandra43,krapbook,Leyvraz2003, FalkovichNature,Falkovich2006}. The aggregating objects may be very different in nature and size, ranging from molecular-scale processes, as aggregating of prions (proteins) in Alzheimer-like diseases \cite{Prions2003}, coagulation of colloids in colloidal solutions (e.g. milk) \cite{Colloid1,Colloid2}, to mesoscopic scale, like aggregation of red blood cells \cite{blood}, or blood clotting \cite{bloodclott}, agglomeration of aerosols in smog \cite{Friedlander,Seinfeld}, to still larger, astrophysical scales, where aggregation of icy particles forms planetary rings \cite{PNAS,constt2} and galaxies form clusters \cite{Galaxies,oort1946gas}. Still, the mathematical description of all such phenomena is similar and based on the celebrated Smoluchowski equations \cite{Smo16,Smo17}. For space homogeneous systems, in the lack of fluxes and sources of particles, they read,
\begin{eqnarray}
\label{loc:input_homo}
\frac{\partial n_k}{\partial t} = \frac{1}{2}\sum_{i+j=k}C_{ij}n_in_j-n_k\sum_{j \geq 1}C_{kj}n_j.
\end{eqnarray}
Here $n_k(t)$ denotes density (a number of objects per unit volume) of clusters of size $k$, that is, clusters comprised of $k$ elementary units -- monomers. $C_{ij}$ are the rate coefficients, which quantify the reaction rates of the cluster merging, $[i]+[j] \to [i+j]$. The first term in the right-hand side of Eq. \eqref{loc:input_homo} describes the increase of the concentration of clusters of size $k$ due to the merging of clusters of size $i$ and $j$ ($1/2$ here prevents double counting). The second term describes the decay of $n_k(t)$ due to the merging of such clusters with all other clusters or monomers.

There exist however plenty of phenomena, where aggregation occurs in non-homogeneous systems with fluxes. Among the prominent examples are sedimentation of coagulating particles \cite{sedim1,sedim2,sedim3}, aggregation of detonation products \cite{detonation1,detonation2,detonation3}, transport of soot emissions in combustion \cite{soot1} and extraterrestrial phenomena with high speed and temperature gradients, like planet formation \cite{proto,proto2}. Here the term ``temperature'' refers to thermodynamic temperature, as well as to granular temperature, associated with the kinetic energy of macroscopic grains; generally, aggregation in space-inhomogeneous systems is of special importance for granular systems, see e.g.  \cite{PNAS,Spahn2006,Longaretti,CanupEsposito,SpahnetalEPL:2004,Das}. Such non-homogeneous systems are much less studied. One can mention the models with one-dimensional advection, \cite{Asymmetric:source,Kirone:source,zagidullin2022aggregation}, where some analytical results have been obtained, and studies, devoted to numerical simulations of spatially non-uniform aggregation \cite{hackbusch2012numerical, bordas2012numerical,chaudhury2014computationally}. In all these studies, a phenomenological generalization of the Smoluchowski equation is used. That is, the standard Smoluchowski equations are simply supplemented by the advection term, yielding the equation,
\begin{equation}
\label{loc:input}
\frac{\partial n_k}{\partial t} + (\vec{u}_k  \cdot \vec {\bf \nabla} ) n_k = \frac{1}{2}\sum_{i+j=k}C_{ij}n_in_j-n_k\sum_{j\geq 1}C_{kj}n_j,
\end{equation}
where $\vec{u}_k$ is the advection velocity of clusters of size $k$. It is  also assumed that the aggregation rates $C_{ij}$ either preserve their form, as for uniform systems, or phenomenological expressions are used, e.g. \cite{Pinsky,FalkovichNature}. To obtain the correct rate coefficients, one needs to derive them from the first principles. For dense systems, it is hardly a solvable problem. It may be done, however, for gases, where microscopic kinetics is described by the Boltzmann equation (BE), e.g.  \cite{krapbook,BrilliantovPoeschelOUP}. There exists a  generalization of the BE for the case of aggregating particles \cite{constt1,constt2,PNAS,Mazza,nature,OsBrilPRE2022}, which may be used to derive Smoluchowski equations, e.g. \cite{constt1,constt2,PNAS} or generalized Smoluchowski equations, e.g. \cite{OsBrilJPA2022,nature,OsBrilPRE2022} with microscopic expressions for the rate coefficients. Here
we use the BE equation to derive Smoluchowski-Euler equations -- new hydrodynamic equations for space-inhomogeneous systems. We observe that Smoluchowski equations not only alter their form, acquiring new terms, but also that  novel kinetic coefficients appear, sharing properties of transport and reaction-rate coefficients.  We perform Direct Simulation Monte Carlo (DSMC), for two representative examples of aggregating systems, and show, that the new theory agrees well with simulation results.

{\em Derivation of Smoluchowski-Euler equations}. The BE  for inhomogeneous systems with aggregation reads,

\begin{eqnarray}
\label{BE}
  \frac{d}{dt} f_k &=& \left( \frac{\partial}{\partial t} + \vec V_k \cdot \frac{\partial}{\partial \vec r} + \frac{{\vec{F}_k}}{m_k}  \cdot \frac{\partial}{\partial \vec V_k} \right) f_k \\
&=& \frac12 \sum\limits_{i+j = k} I_{ij}^{{\rm agg},1}  - \sum\limits_{j = 1}^{\infty} I_{kj}^{{\rm agg},2}   + \sum\limits_{j = 1}^{\infty} I_{kj}^{{\rm res}} , \nonumber
\end{eqnarray}
where $f_k=f_k(\vec{V}_k, \vec{r}, t)$ is the velocity distribution function (VDF) for clusters of size $k$ and velocity $\vec V_k$ at a point $\vec{r}$ and time $t$ and $\vec{F}_k$ is the external force.
The aggregation collision integrals have the following form \cite{nature,OsBrilPRE2022}:
\begin{eqnarray}
I_{ij}^{\rm agg,1} &= &\sigma^2_{ij}\int d\vec{V}_id\vec{V}_jd\vec{e}|\vec{V}_{ij}\cdot \vec{e}| \Theta(-\vec{V}_{ij}\cdot \vec{e})\Theta(W_{ij}-E_{ij}) \nonumber \\
&\times& \delta \left(M_{ij}\vec{V}_{i+j}-m_i\vec{V}_i-m_j\vec{V}_j \right) f_if_j\\
I_{ij}^{\rm agg,2} \!&= &\!\sigma^2_{ij}\!\!\int \!\!\!d\vec{V}_jd\vec{e}|\vec{V}_{ij}\!\!\cdot \vec{e}| \Theta(\!-\vec{V}_{ij}\cdot \vec{e})\Theta(W_{ij}\!-\!E_{ij})f_if_j. \nonumber
\end{eqnarray}
Here $\sigma_{ij}=(\sigma_i+\sigma_j)/2 $ is the collision cross-section ($\sigma_i$ is the diameter of a cluster of size $i$). $W_{ij}$ quantifies the energy of the attractive (adhesion) barrier. If the relative kinetic energy of colliding particles, of size $i$ and $j$, at the end of a collision, $E_{ij}=\varepsilon^2 \mu_{ij}V_{ij}^2/2 $, exceeds $W_{ij}$, the particles bounce, otherwise, they merge. Here $\mu_{ij}=m_im_j/M_{ij}$ is the reduced mass, $M_{ij}=m_i+m_j$ and $\vec{V}_{ij}=\vec{V}_i-\vec{V}_j$ is the relative velocity. $\varepsilon$ is the restitution coefficient \cite{BrilliantovPoeschelOUP}, which we assume to be constant and $\sigma_{ij}^2|\vec{V}_{ij}\cdot \vec{e}|$, where $\vec{e}$ is the unit vector, joining particles' centers at the collision instant,  gives the volume of the collision cylinder. Finally, $\Theta(-\vec{V}_{ij}\cdot \vec{e})$,  selects only approaching particles \cite{krapbook,BrilliantovPoeschelOUP}. The factor with $\delta$-function in the integrand of $I_{ij}^{\rm agg,1}$, guarantees the momentum conservation at merging. The restitution integral,  $I_{ij}^{\rm res}$ has the conventional form for bouncing collisions, see e.g. \cite{krapbook,BrilliantovPoeschelOUP}, but contains an additional factor in the integrand, $\Theta(E_{ij}-W_{ij})$, which guarantees that the corresponding collisions are bouncing. Here we do not need an explicit form for this quantity; it is presented in the Supplemental Material (SM) \cite{Note2}. Generally, Eq. \eqref{BE} may contain a source of monomers  or clusters, ${\cal J}_k ( \vec V_k )$ \cite{Leyvraz2003,Sire, Colm_Kolmogor,Colm_Oscill}.

Generally, BE \eqref{BE} is not solvable. Fortunately, for most practical applications only fluid behavior at \textit{hydrodynamic} stage of evolution is important. The initial conditions are forgotten at this stage and the dependence on time and space of the velocity distribution function (VDF) occurs only trough  hydrodynamic fields, which are the first few moments of the VDF  \cite{Resibua,Garzo}. The VDF itself is approximated by a function with the same few first moments as a true one. The most simple  VDF is,
\begin{equation}
\label{eq:Max}
  f_k \left( \vec V_k, \vec{r},t \right) = \frac{n_k }{\left( 2 \pi \theta_k \right)^{3/2}} e^{\frac{-\left( \vec V_k - \vec u_k \right)^2}{2 \theta_k}},
\end{equation}
which is Maxwellian, with five moments -- zero, first and second-order. These are:  $n_k = n_k\left(\vec{r},t \right) = \int \!f_k d \vec{V}_k$  -- the number density of clusters  of size $k$,
$\vec{u}_k=\vec{u}_k\left( \vec{r},t \right) = n_k^{-1} \int \vec{V}_k f_k d \vec{V}_k$ -- the respective flux velocity and  $\theta_k=T_k\left( \vec{r},t \right)/m_k$ -- the reduced  temperature of such clusters, $\theta_k = \tfrac13 n_k^{-1} \int f_k  (\vec{V}_k -\vec{u}_k)^2 f_k d \vec{V}_k$. The next approximation is Grad's 13-moment approach \cite{Grad} (or 14-moment for granular gases \cite{Garzo}), which describes deviations of  VDF from the Maxwellian, see SM. Here we address the hydrodynamic evolution stage with Eq. \eqref{eq:Max} for the  VDF,  since for aggregating systems it is close to the Maxwellian, see SM. This yields a relatively simple theory, with an acceptable accuracy. It corresponds to Euler's hydrodynamics, e.g. \cite{Resibua}. The Navier-Stokes's level of description with 13 (or 14) moments \cite{Resibua,Garzo} may also be elaborated, but it is rather   complicated, see SM for detail.

Integrating BE \eqref{BE} over the velocities $\vec{V}_k$
yields,
\begin{equation}
\label{eq:n}
  \frac{\partial}{\partial t} n_k \!+ \!\vec \nabla \cdot \left( n_k \vec u_k \right) \!= \!\frac{1}{2} \!\sum\limits_{i+j=k} \!C_{ij} n_i n_j - \!\sum\limits_{j = 1}^{\infty} \!C_{kj} n_k n_j \equiv S_1^{(k)}.
\end{equation}
 These are Smoluchowski equations for inhomogeneous systems. Microscopic expressions for $C_{ij}$ may be obtained for an arbitrary aggregation barrier $W_{ij}$, see SM.  It is instructive, however, to consider a simpler, but still very important case of $W_{ij}/T_k \gg 1$ for all $k$, when practically all the collisions are merging. In this case, one can neglect restitution collisions, $I_{ij}^{\rm res}=0$, and  all expressions are significantly simplified. The rate coefficients, corresponding to VDF \eqref {eq:Max}, read (see SM for derivation detail):
\begin{equation}\label{eq:cagg}
   C_{ij} = \sqrt{2 \pi} \sigma_{ij}^2 \sqrt{\theta_i + \theta_j} \left[e^{-c^2/4} + \frac{\sqrt{\pi} (c^2+2)}{2c}{\rm erf}(\tfrac{c}{2}) \right] ,
\end{equation}
where $c\equiv \sqrt{2/(\theta_i + \theta_j)} |\vec{u}_i- \vec{u}_j|$. Note that in the limiting case of vanishing fluxes, $\vec{u}_i = \vec{u}_j \to 0$, the rate coefficient reduces to the known one, $C_{ij}^{(0)} = 2 \sqrt{2 \pi} \sigma_{ij}^2 \sqrt{\theta_i + \theta_j}$ \cite{OsBrilJPA2022,nature,OsBrilPRE2022}. For the other limiting case   $\left|\vec{u}_i - \vec{u}_j \right| \to \infty$, the rate coefficients take the form $C_{ij}^{(1)} = \pi \sigma_{ij}^2|\vec{u}_i - \vec{u}_j|$, used in the sedimentation problem \cite{FalkovichNature,zagidullin2022aggregation}. In Ref. \cite{proto} the authors utilized the phenomenological kernel, $C_{ij} = C_{ij}^{(0)}+C_{ij}^{(1)}$. Obviously, in the general case, the reaction rate kernel noticeably differs from the phenomenological one.

Note that neither $\vec{u}_k$ nor $T_k$ are independent variables -- they  evolve subject to the aggregation kinetics. To obtain equations for these quantities, we multiply the BE \eqref{BE} with $\vec{V}_k$ and $\tfrac13 (\vec V_k - \vec u_k)^2$ and integrate over $\vec{V}_k$. This yields the equations for $\vec{u}_k$,  and $\theta_k=T_k/m_k$:
\begin{eqnarray}
   \frac{\partial}{\partial t} \left( n_k \vec u_k \right) &+&  n_k \vec u_k \cdot \vec \nabla \vec u_k + \vec u_k \vec \nabla \cdot \left( n_k \vec u_k \right)
  + \vec \nabla \left( n_k \theta_k \right) \nonumber \\ - n_k \vec{F}_k/m_k
 &=& \frac{1}{2} \sum\limits_{i+j=k} \!\vec P_{ij} n_i n_j \!-\! \sum\limits_{j = 1}^{\infty} \! \vec R_{kj} n_k n_j \equiv \vec{S}_2^{(k)},   \label{eq:u} \\
   \frac{\partial}{\partial t} \left( n_k \theta_k \right) &+&  n_k \vec u_k \cdot \vec \nabla  \theta_k +  \theta_k \vec \nabla \cdot \left( n_k \vec u_k \right) + \frac{2}{3} n_k \theta_k \vec \nabla \cdot \vec u_k \nonumber \\
  & =& \frac{1}{2} \sum\limits_{i+j=k} \!B_{ij} n_i n_j \!-\! \sum\limits_{j = 1}^{\infty}\!D_{kj} n_k n_j \equiv S_3^{(k)}.  \label{eq:t}
\end{eqnarray}
While terms on the left-hand side of Eqs. \eqref{eq:u} and \eqref{eq:t} have the same form as for conventional Euler equations (with pressure $p=n_k\theta_k m_k)$, in the right-hand-side there appear novel kinetic coefficients -- vectorial  $\vec{P}_{ij}$, $\vec{R}_{ij}$, and scalar coefficients $B_{ij}$, $D_{ij}$, which may be dubbed, respectively, as ``flux-reaction'' and ``energy-reaction'' rates. They depend on fluxes $\vec{u}_k $ and temperatures $T_k$.

Hence, Eqs. \eqref{eq:n}, \eqref{eq:u} and \eqref{eq:t} form a closed set of equations for $n_k(\vec{r},t)$, $\vec{u}_k (\vec{r},t)$ and $\theta_k =T_k(\vec{r},t)/m_k$ and may be called {\em Smoluchowski-Euler } equations -- the first-principle hydrodynamic equations for aggregating non-uniform systems.  Referring to the derivation detail to SM, we present here the expressions for the novel kinetic coefficients for the case, when all collisions are merging and VDF is Maxwellian:
\begin{equation}\label{PRij}
\begin{aligned}
& \vec P_{ij} = C_{ij} \vec \mu + 2 \sqrt{2 \pi} \sigma_{ij}^2  q \vec c \; \times \\
& \times \left( e^{-c^2/4} \left( 1/c^2 + 6 \right) + \frac{\sqrt{\pi}}{2c} \left( 1 - 2/c^2 \right) {\rm erf} \left( c/2 \right) \right), \\
& \vec R_{ij} = C_{ij} \vec u_i + 2\sqrt{2 \pi} \sigma_{ij}^2 \theta_i \vec c \; \times \\
& \times \left( e^{-c^2/4} \left( 1/c^2 + 6 \right) + \frac{\sqrt{\pi}}{2c} \left( 1 - 2/c^2 \right) {\rm erf} \left( c/2 \right) \right),
\end{aligned}
\end{equation}
where $\vec \mu = (m_i \vec u_i + m_j \vec u_j)/M_{ij}$, $\vec{c} = \sqrt{2}(\vec u_i -\vec u_j)/\sqrt{\theta_i+\theta_j}$ and $q = (T_i - T_j)/M_{ij}$. We also present the expression for  scalar coefficients, $D_{ij}$, referring to SM for the expression for $B_{ij}$, which are too cumbersome to be given here.
\begin{eqnarray}
\label{eq:Dij}
  D_{ij}  &=& C_{ij} \frac{\theta_i \theta_j}{\theta_i + \theta_j} + \\
  &+&  \frac{\sqrt{2 \pi} \sigma_{ij}^2 \theta_i^2}{\sqrt{\theta_i + \theta_j}} \left[ e^{-c^2/4} + \frac{\sqrt{\pi}}{6c} \left( 10 + 3c^2 \right) {\rm erf} \left( \tfrac{c}{2} \right) \right]. \nonumber
\end{eqnarray}

Note that terms responsible for viscosity and thermal conductivity are neglected in Smoluchowski-Euler equations, as compared to the terms describing aggregation; the contribution of such terms is discussed in SM.

Now we consider some representative applications of new hydrodynamic  equations and demonstrate a good agreement of the new theory with the DSMC results.

{\em Dust sedimentation with aggregation. }
Consider now aggregation kinetic of vertically falling particles (e.g. soot) from a source, when  their horizontal motion may be neglected. That is, we assume that the system is homogeneous in $x$ and $y$ directions and non-zero flux exists only in the vertical, $z$-direction, $u_k = u_{z,k}$. We also assume that the particles are massive enough, so that the thermal speed, gained from collisions with the molecules of the surrounding gas, is negligible (see the discussion in SM). This implies $\theta_k = \left\langle \left(v_{z,k} - u_k \right)^2 \right\rangle$. The particles experience the gravitational acceleration $g$ and are  slowed down by the atmosphere.  Here  we use the Stokes relation for the viscous friction force, $\vec{F}=-3\pi \eta \sigma \vec{V}$, where $\eta$ is the gas viscosity, $\sigma$ and $\vec{V}$ are respectively particles' diameter and velocity; this implies a steady velocity,  $V_{\rm eq}= mg/(3 \pi \eta \sigma)$. The kinetic equations, describing this quasi-one dimensional system for the case of all-merging collisions read,  see SM for detail:
\begin{align}
 & \frac{\partial}{{\partial t}}{n_k} + \frac{\partial}{\partial z} \left( {n_k  u_k} \right) = S_1^{(k)} +J\delta(z) \left( \delta_{k1} + \delta_{k2} \right), \label{eq:eulercz} \\
 & \frac{\partial}{{\partial t}} \left( {n_k}{u_k} \right) + \frac{\partial}{\partial z} \left( n_k u_k^2 \right) + \frac{\partial}{\partial z} \left( {n_k \theta_k } \right) + n_k g  \nonumber \\
 &+3 \pi \eta \frac{\sigma_k}{m_k} n_k u_k =S_{2,z}^{(k)} - J \delta(z) \left( \delta_{k1} + 2^{2/3} \delta_{k2} \right) , \label{eq:eulervz} \\
 & \frac{\partial}{{\partial t}} \left({n_k}{\theta_k} \right) + \frac{\partial}{\partial z} \left( n_k u_k \theta_k \right) + 2 n_k \theta_k \frac{\partial}{\partial z} u_k \\
 & + 6 \pi \eta \frac{\sigma_k}{m_k} n_k \theta_k
 = S_3^{(k)} . \label{eq:eulerez}
\end{align}
Here we use the source of monomers and dimers, located at $z=0$, equal to
$
{\cal J} = J\delta(z)[\delta_{k,1}(V_1-  V_{\rm eq,1}) +\delta_{k,2}( V_2- V_{\rm eq,2})]
$. These particles have the corresponding steady velocities, and we apply the appropriate units for them (see below). Note that the monomers cannot aggregate with themselves, having the same steady velocity with any variance quickly damped by air friction. Thus, we need to consider at least two different sizes. $S_1$, $\vec S_2$ and $S_3$ have been defined above, in Eqs. \eqref{eq:n}, \eqref{eq:u} and \eqref{eq:t}, however, the kinetic coefficients, $C^{(1D)}_{ij}$, $B^{(1D)}_{ij}$, $D^{(1D)}_{ij}$ and $\vec P^{(1D)}_{ij}$, $\vec R^{(1D)}_{ij}$  are now different there, since they describe quasi-one dimensional case. The derivation and structure of these quasi-1D coefficients are very similar to the 3D case and are detailed in SM. Here we present $C^{(1D)}_{ij}$, referring to SM for other coefficients:
\[
  C^{(1D)}_{ij} = \sqrt{2 \pi} \sigma_{ij}^2 \sqrt{\theta_i + \theta_j} e^{-c^2/4} + \pi \sigma_{ij}^2 \left| u_i - u_j \right| {\rm erf} \left( c/2 \right).
\]
Note that the reduced temperatures, $\theta_k$, are non-zero, since large  particles  can originate in many different ways, from particles falling with significantly different velocities. This creates a noticeable velocity variance (reduced temperature) for each particle size. Still, the variance cannot infinitely grow, as it is quickly dumped by air friction; any  horizontal speed is also quickly dumped.

In computations, we choose the physical units with the unit diameter of monomers, ($\sigma_1=1$, $\sigma_k=k^{1/3}$), unit mass of monomers ($m_1=1$, $m_k = k$)
\cite{Note1} and unit equilibrium velocity of monomers, $V_{1, \rm eq}=m_1 g/(3\pi \eta  \sigma_1)= g/(3 \pi \eta)=1$, which yields $\eta =g/(3\pi)$. In these units, the equilibrium velocity of $k$-mers (in the absence of aggregation) reads, $V_{k, \rm eq}=k^{2/3}$. The relevant characteristic length of the system is $l_0 = \sigma_1$, the characteristic time is $\tau_0 = \sigma_1 / V_{\rm 1,eq}$. In these units, the system is characterized by two dimensionless parameters --  the dimensionless gravity, $g^* = g \tau_0^2 / l_0 = (36 \pi^2 \eta^2 \sigma_1^3)/(m_1^2 g)$ and the total source intensity (number of particles per unit time per unit area) $J^* = 2 J \tau_0 l_0^2 = 2 n_0 \sigma_1^3$ for $J = n_0 V_{\rm 1,eq}$, where $n_0$ is the number density of monomers (we assume equal number of monomers and dimers with equilibrium speeds, yielding the coefficient 2).
\begin{figure}[!htb]
\centering
\begin{subfigure}[t]{0.49\linewidth}
\caption{}\label{subfig:a}
\includegraphics[width=\columnwidth]{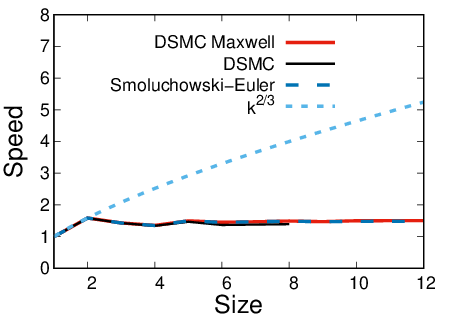}
\end{subfigure}
\begin{subfigure}[t]{0.49\linewidth}
\caption{}\label{subfig:b}
\includegraphics[width=\linewidth]{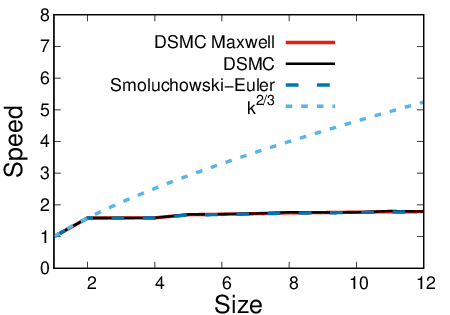}
\end{subfigure}
\begin{subfigure}[t]{0.49\linewidth}
\caption{}\label{subfig:c}
\includegraphics[width=\linewidth]{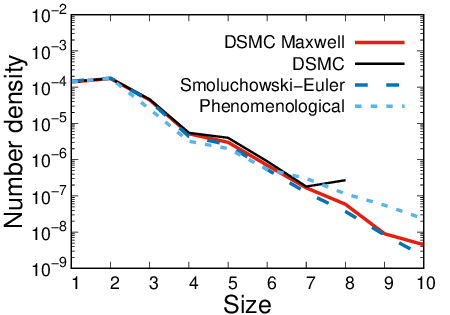}
\end{subfigure}
\begin{subfigure}[t]{0.49\linewidth}
\caption{}\label{subfig:d}
\includegraphics[width=\linewidth]{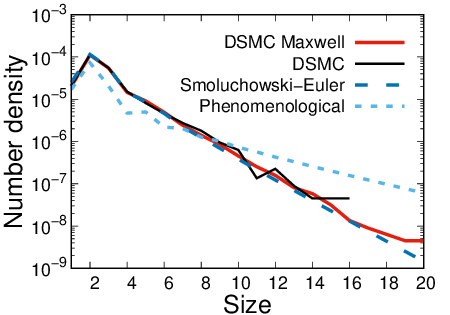}
\end{subfigure}
\caption{\textbf{(a)}, \textbf{(b)}: Partial flow velocities $u_k$ of particles of size $k$, 
at $z = -50 \, m$ (a) and at $z = -250 \, m$ (b). The steady-state velocities $V_{k, \rm eq} = k^{2/3}$ ($V_{\rm 1,eq}=1 $) are shown for the approximation when aggregation is neglected. \textbf{(c)}, \textbf{(d)}: Comparison of the particle size distribution for the new theory, Eqs. \eqref{eq:eulercz}-\eqref{eq:eulerez} (long dashes), phenomenological Eqs. \eqref{loc:input}, with fixed speeds, $u_k=k^{2/3}$, directed downward, and kinetic rates, $C_{ij}=\pi \sigma_{ij}^2|u_i-u_j|$, (short dashes) and DSMC results (solid lines) for $z = -50 \, m$ (c) and  $z = -250 \, m$ (d). We use 100 vertical layers and about $5 \cdot 10^3$ particles per layer for standard DSMC and $5 \cdot 10^4$ particles per layer for DSMC with Maxwell assumption for VDF.  The dimensionless monomer source and gravity are $J^*=4.5 \times 10^{-4}$ and $g^*=4.0 \times 10^{-4}$. The source of monomers and dimers is located at $z=0$.
}
\label{fig:fall}
\end{figure}
We solve Eqs. \eqref{eq:eulercz}-\eqref{eq:eulerez} numerically and perform two types of DSMC -- the standard one and the simplified, with the assumption of Maxwellian VDF, see SM for detail. We use $J^*=4.5 \times 10^{-4}$ and $g^*= 4.0 \times 10^{-4}$, which may correspond, e.g., to the following parameters of soot particles: diameter $\sigma_1=1.35\times 10^{-3}\, m$,  monomer mass $m_1 = 2.7 \times 10^{-6} \, kg$, cluster mass density $\rho=2\times 10^3\, kg/m^3$, number density $n_0=9.2 \times 10^4 \, m^{-3}$, flow speed $V_{\rm 1,eq}=12\,m/s$, for the air viscosity $\eta=1.8\times 10^{-5}\, kg/m/s$ and $g=9.8\, m/s^2$.

In Fig. \ref{fig:fall} we compare the solution of the Smoluchowski-Euler equations,  \eqref{eq:eulercz}-\eqref{eq:eulerez} and of the phenomenological equation \eqref{loc:input}, where the steady-state speeds, $V_{k, \rm eq}=k^{2/3}$ are used for the flux velocities, $u_k^{\rm (ph)}=k^{2/3}$; here we also show DSMC results. As may be seen from the figure, the actual flux velocity $u_k$ significantly differs from its phenomenological approximation, $u_k^{\rm (ph)}=k^{2/3}$. Furthermore, the size distributions of the aggregates are also very different, especially for large clusters,  where they differ by the orders of magnitude (note the logarithmic scale for $n_k$). Hence, while the first-principle theory agrees well with the DSMC results, the phenomenological equations cannot provide a reliable description of the processes. One can also see, that the Maxwell distribution assumption does not affect the DSMC results.

{\em Explosion with aggregation.}  Another important example is the aggregation kinetics in a system of particles (debris), emerging in an explosion in a vacuum, with the center at $r=0$. We consider the spherically symmetric case and neglect all components of the particles' velocities except the radial one. That is, we assume that $u_k=u_{r,k}$ and similarly, $\theta_k=\langle (V_{r,k}-u_k)^2 \rangle$. In the lack of gravity, the governing equations read:
\begin{align}
 & \frac{\partial}{{\partial t}}{n_k} + \frac{1}{r^2} \frac{\partial}{\partial r} \left( r^2 {n_k  u_k} \right) = S_1^{(k)}, \label{eq:eulercr} \\
 & \frac{\partial}{{\partial t}} \left( {n_k}{u_k} \right) + \frac{1}{r^2} \frac{\partial}{\partial r} \left( r^2 n_k\left(   u_k^2  +  \theta_k \right) \right) = S_{2,r}^{(k)}, \label{eq:eulervr} \\
 & \frac{\partial}{{\partial t}} \left({n_k}{\theta_k} \right) + \frac{1}{r^2}\frac{\partial}{\partial r} \left( r^2  n_k u_k \theta_k \right) + 2 n_k \theta_k \frac{\partial}{\partial r} u_k = S_3^{(k)}. \label{eq:eulerthetr}
\end{align}
In $S_1$, $\vec S_2$ and $S_3$ we use the same kinetic coefficients as for the quasi-one dimensional case discussed above. For the initial conditions, we assume that the number density of monomers at $r=r_0=1$ is $n_1(r_0)=1$.  The monomers have Maxwell distribution of  the initial radial velocities, with the average velocity $u_0 = 1$ and variance $\theta_0 = 0.2$.
\begin{figure}[htb]
\centering
\begin{subfigure}[t]{0.49\linewidth}
\caption{}\label{subfig2:b}
\includegraphics[width=\linewidth]{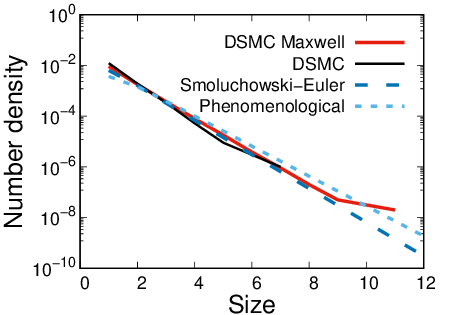}
\end{subfigure}
\begin{subfigure}[t]{0.49\linewidth}
\caption{}\label{subfig2:c}
\includegraphics[width=\linewidth]{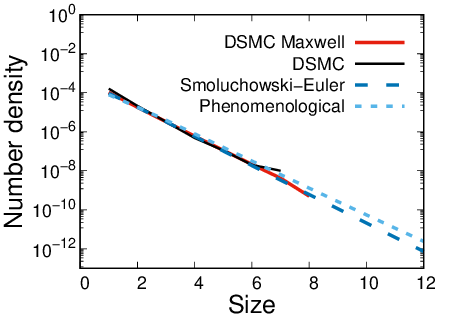}
\end{subfigure}
\caption{
The size distribution of the aggregates at time $t=8$ for the distances from the epicenter of  $r = 6$ (a) and $r=10$ (b).
The results for the phenomenological model, Eqs. \eqref{loc:input} and \eqref{eq:Eulu} (short dashes) are compared with the solution of the Smoluchowski-Euler equations \eqref{eq:eulercr}-\eqref{eq:eulerthetr} (long dashes) and DSMC results (solid lines)  with $10^6$ initial particles for standard DSMS and $10^8$ particles for DSMC with the Maxwellian assumption for VDF.  At $t = 0$ only monomers with the average radial velocity $u_0 = 1$ and variance $\theta_0 = 0.2$ were at $r = 1$.
%
}
\label{fig:radial}
\end{figure}

In Fig. \ref{fig:radial} we present the size distribution of particles at different radial distances, $r=6$ and $r=10$, at time $t=8$ after the explosion. Here we compare the DSMC results, the solution of the Smoluchowski-Euler equations, \eqref{eq:eulercr}-\eqref{eq:eulerthetr},  and the solution of the  phenomenological equations \eqref{loc:input}, supplemented by Euler equations for flow velocities,
\begin{equation}
\label{eq:Eulu}
  \frac{\partial u_k}{\partial t}  + u_k \frac{\partial u_k}{\partial r}  + \frac{1}{n_k r^2} \frac{\partial}{\partial r} \left(  r^2 n_k \theta_k \right) = 0,
\end{equation}
and fixed temperatures $T_k = k \theta_0 = \mathrm{const}$.
As may be seen from the figure, the size distribution, obtained from Eqs. \eqref{eq:eulercr}-\eqref{eq:eulerthetr} of the new theory agrees well with the DSMC, while it significantly differs for the phenomenological theory, especially for large aggregates. The total number density, $N=\sum_k n_k$,  differs not so much -- the observed difference was about  $20\%$. We expect, however, that in the course of time, the predictions of the phenomenological theory will deviate more and more from the results of the DSMC and the first-principle theory.

{\em Conclusion.} We report new hydrodynamic equations -- Smoluchowski-Euler equations, which describe aggregation kinetics in space-inhomogeneous fluids with fluxes; aggregation in granular systems is the most prominent example of such phenomena.  We derive these equations for the number density of aggregates, their average velocity and kinetic temperature from the first principles, starting from the Boltzmann equation, and obtain microscopic expression for the aggregation rate coefficients. These coefficients significantly differ from the respective coefficients for homogeneous systems without fluxes and from their phenomenological generalization. Surprisingly, we reveal, that apart from the conventional aggregation-rate coefficients for the cluster densities,  a set of new kinetic coefficients appears in the equations for flux velocities and temperatures. We obtain microscopic expressions for the new kinetic coefficients, which share properties of transport and reaction-rates coefficients. We consider two representative examples of the application of the new equations --  the sedimentation of aggregating particles and aggregation of particles in an explosion and perform Direct Simulation Monte Carlo (DSMC) for these systems. We demonstrate that predictions of the new theory agree well with the DSMC results, but significantly differ from the results of the currently used phenomenological theory. This indicates that the phenomenological description of aggregation processes in non-homogeneous fluids with fluxes is not reliable, and that one needs to apply new, first-principle Smoluchowski-Euler equations, reported here.  Since the new theory is based on the Boltzmann equations, its application is limited to dilute systems (gases), including dilute granular systems.  Derivation of the Smoluchowski-Euler equations for dense media remains challenging.  In our work we did not consider cluster fragmentation -- the process opposite to aggregation; it may be important for some systems, e.g. for planetary rings. Although the implementation of disruptive collisions  is straightforward,  it requires a microscopic fragmentation model, which will be addressed in future studies. %

\bibliographystyle{unsrt}

\appendix

\begin{widetext}

\begin{center}
\Large \textbf{Supplemental Material  for \\
Hydrodynamic equations for space-inhomogeneous aggregating \\ fluids with first-principle
kinetic coefficients}
\end{center}

Here we present some valuable information and detailed derivation of the relations of the main text and discuss 14-moment Grad's approach. Namely, we start from the microscopic Boltzmann equation and demonstrate how the first-principle hydrodynamic equations -- Smoluchowski-Euler equations, may be obtained from the Boltzmann equation. Here we also present the derivation of the microscopic expressions of new kinetic coefficients, for 3D and quasi-1D systems. We also demonstrate how next-order theory (14-moment Grad's approximation), with non-Maxwellian velocity distribution function may be elaborated. Based on the numerical experiment, we show that the velocity distribution function in aggregating systems is very close to the Maxwellian. Finally, we discuss the details of the numerical scheme used to solve the kinetic equations.



\section{Introduction}
Aggregation kinetics in space inhomogeneous systems is observed in many natural systems and can take place  in industrial processes. The former systems may be exemplified by numerous phenomena in living systems, by atmospheric or astrophysical phenomena, and the latter -- by coagulation processes in food, pharmaceutic or building industries. Hence an adequate modelling of such processes is of primary importance. The main tool to model the aggregation kinetics is the phenomenological Smoluchowski equation, which was first formulated for space-uniform systems. This equation was  phenomenologically modified  to account for fluxes and space inhomogeneity. Still, it is desirable to obtain kinetic equations derived from the first principles. This may be done using the microscopic Boltzmann equation.

\section{Boltzmann equation}
The  Boltzmann equation  still  remains one of the main pillars of non-equilibrium statistical mechanics and is exploited in many areas of kinetic theory, ranging from classical gas dynamics and dynamics of granular gases \cite{BrilliantovPoeschelOUP,Garzo,ChapmanCowling:1970}, to  aggregation and fragmentation phenomena \cite{SpahnetalEPL:2004,CanupEsposito,Longaretti,Spahn2006,esposito2006}, traffic  and active matter modelling, e.g. \cite{krapbook,BEacpa}. The aggregation kinetics described by the Boltzmann equation refers to atmospheric phenomena, such as coagulation of dust or airborne particles, e.g.  \cite{Hidy,Drake,Srivastava1982,Friedlander}, the behavior of astrophysical systems -- planetary rings and  interstellar dust clouds, e.g. \cite{PNAS,constt2,Galaxies,oort1946gas}. The complete description of aggregation is very complicated.  Therefore, several simplifying assumptions are applied. First, it is assumed that each aggregate may be completely characterized by its mass, which is determined by the number of elementary units (monomers) comprising the cluster. Second, in the realm of the Boltzmann approach, it is assumed that it is also characterized by velocity. Hence, the shape of an aggregate and its angular motion are ignored; under these assumptions, one can formulate Boltzmann-like  equations for aggregating particles.

Consider the case, when the system can be approximated by discretely sized clusters with masses $m_i = im_1$, with $m_1=1$. Then the velocity distributions $f_k \left( \vec V_k , \vec r,t \right)$ for each cluster of size $k$ obey the following system of Boltzmann equations (BEs),

\begin{equation}\label{eq:boltz}
  \frac{d}{dt} f_k \left( \vec V_k \right) = I^{\rm agg}-I^{\rm res} =
  \frac12  \sum\limits_{i+j = k} I_{ij}^{{\rm agg},1} \left( \vec V_k \right) - \sum\limits_{j = 1}^{\infty} I_{kj}^{{\rm agg},2} \left( \vec V_k \right) - \sum\limits_{j = 1}^{\infty} I_{kj}^{{\rm res}, 1} \left( \vec V_k \right) + \sum\limits_{j = 1}^{\infty} I_{kj}^{{\rm res},2} \left( \vec V_k \right) + {\cal J}_k \left( \vec V_k \right),
\end{equation}
where we skip the dependence on $\vec r$ and $t$ and ${\cal J}_k\left( \vec V_k \right)$ describes the source of particles of size $k$ and velocity $\vec V_k $. We also have the following Boltzmann collision integrals which account for aggregative and restitutive collisions:
\begin{equation}\label{eq:bolint}
\begin{aligned}
  I_{ij}^{{\rm agg},1} \left( \vec V_{i+j} \right) & = \sigma_{ij}^2 \iiint \left| \left( \vec V_i - \vec V_j \right) \cdot \vec e \right| \Theta \left( -\left( \vec V_i - \vec V_j \right) \cdot \vec e \right) \Theta \left( W_{ij} - \frac{ \varepsilon^2 m_i m_j \left| \vec V_i - \vec V_j \right|^2}{2\left( m_i + m_j \right)} \right) \\
  & \times \delta \left( \vec V_{i+j} - \frac{m_i \vec V_i + m_j \vec V_j}{m_i + m_j} \right) f_i \left( \vec V_i \right) f_j \left( \vec V_j \right) \, d \vec V_i \, d \vec V_j \, d \vec e,
        \end{aligned}
\end{equation}
\begin{equation}\label{eq:bolint1}
\begin{aligned} \nonumber
  I_{ij}^{{\rm agg},2} \left( \vec V_i \right) & = \sigma_{ij}^2 \iint \left| \left( \vec V_i - \vec V_j \right) \cdot \vec e \right| \Theta \left( -\left( \vec V_i - \vec V_j \right) \cdot \vec e \right) \Theta \left( W_{ij} - \frac{\varepsilon^2 m_i m_j \left| \vec V_i - \vec V_j \right|^2}{2\left( m_i + m_j \right)} \right) \\
  & \times f_i \left( \vec V_i \right) f_j \left( \vec V_j \right) \, d \vec V_j \, d \vec e, \\
  I_{ij}^{{\rm res},1} \left( \vec V_i \right) & = \sigma_{ij}^2 \iiint \left| \left( \vec V_i - \vec V_j \right) \cdot \vec e \right| \Theta \left( -\left( \vec V_i - \vec V_j \right) \cdot \vec e \right) \Theta \left( \frac{ \varepsilon^2 m_i m_j \left| \vec V_i - \vec V_j \right|^2}{2\left( m_i + m_j \right)} - W_{ij} \right) \\
  & \times f_i \left( \vec V_i \right) f_j \left( \vec V_j \right) \, d \vec V_j \, d \vec e, \\
  I_{ij}^{{\rm res},2} \left( \vec V_i \right) & = \sigma_{ij}^2 \iiint \left| \left( \vec V_i - \vec V_j \right) \cdot \vec e \right| \Theta \left( -\left( \vec V_i - \vec V_j \right) \cdot \vec e \right) \Theta \left( \frac{ \varepsilon^2 m_i m_j \left| \vec V_i - \vec V_j \right|^2}{2\left( m_i + m_j \right)} - W_{ij} \right) \\
  & \times \frac{1}{\varepsilon^2} f_i \left( \vec V_i - \frac{1 + \varepsilon}{2 \varepsilon} \cdot \frac{m_j}{m_i + m_j} \left( \left( \vec V_i - \vec V_j \right) \cdot \vec e \right) \vec e \right) \\
  & \times f_j \left( \vec V_j + \frac{1 + \varepsilon}{2 \varepsilon} \cdot \frac{m_i}{m_i+m_j} \left( \left( \vec V_i - \vec V_j \right) \cdot \vec e \right) \vec e \right) d \vec V_j \, d \vec e.
\end{aligned}
\end{equation}
\end{widetext}
Here we introduce the radius of the collision cylinder $\sigma_{ij}$ and the potential barrier $W_{ij}$, which can be used to determine, whether a collision leads to aggregation or bouncing \cite{SpahnetalEPL:2004,constt1,PNAS,nature}; $\Theta(x)$ in the above equations are the unit Heaviside  step functions. These functions in the integrands select only approaching particles and distinguish between bouncing and aggregating collisions for the particular aggregation barrier  $W_{ij}$. That is, if the relative kinetic energy at the very end of a collision, $\varepsilon^2 m_im_j(\vec V_i - \vec V_j)^2/2(m_i+m_j)$,  is smaller than the potential barrier energy, the particles merge. Otherwise, they bounce off. The unit vector $\vec e$ specifies the collision geometry -- it is directed along the inter-particle centers at the collision instant. Finally, $\vec V_i$ and $\vec V_j$ denote the velocities of the colliding particles.

The first aggregation integral describes the rate of change of the distribution function for clusters of size $k = i + j$, emerging in the collisions of clusters of size $i$ and $j$. The second one quantifies the rate of change of the number of clusters of size $i$ with velocities $\vec V_i$ which disappear in aggregative collisions. The restitution integral $I_{ij}^{{\rm res},1}$ accounts for the number of clusters that change their velocity  from $\vec V_i$ to $\vec V_i^{\prime}$ after the collision,
$$
\vec{V}_{i/j}^{\prime} = \vec{V}_{i/j}  \mp \frac{1 + \varepsilon}{2 } \cdot \frac{m_i}{(m_i+m_j)} \left( \vec{V}_{ij} \cdot \vec e \right) \vec e.
$$
$I_{ij}^{{\rm res},2}$ describes, respectively, the ``inverse'' collisions with the velocities
$$
\vec{V}_{i/j}^{\prime \prime} = \vec{V}_{i/j}  \mp \frac{1 + \varepsilon}{2 \varepsilon} \cdot \frac{m_i}{(m_i+m_j)} \left( \vec{V}_{ij} \cdot \vec e \right) \vec e,
$$
which end up with the velocities $\vec{V}_{i/j}$.  The length of the collision cylinder for the inverse collisions is rescaled accordingly by a factor of $1/\varepsilon^2$,   see \cite{BrilliantovPoeschelOUP} for more detail.

The full time derivative for each size $k$ can be written as follows:
\begin{multline}
\label{eq:dfdt}
  \frac{d}{dt} f_k \left( \vec V_k, \vec r, t \right) = \\
  = \left( \frac{\partial}{\partial t} + \vec V_k \cdot \frac{\partial}{\partial \vec r} + \frac{\vec F_k}{m_k} \cdot \frac{\partial}{\partial \vec V_k} \right) f_k \left( \vec V_k, \vec r, t \right),
\end{multline}
where $\vec F_k$ is the external force. Here we explicitly state that the velocity distribution can vary in time and space.

The source of particles used in the main text (monomers and dimers added at $z=0$ with their equilibrium sedimentation velocities) reads:
$$
{\cal J} (\vec V_k) = J\delta(z)\left[\delta_{k,1}(\vec V_1- \vec V_{\rm eq,1}) +\delta_{k,2}(\vec V_2- \vec V_{\rm eq,2}) \right].
$$

Our extensive Monte Carlo simulations have demonstrated that for an aggregating system, the velocity distribution function for all clusters is very close to Maxwellian, except for a small region for clusters with almost vanishing velocity, see  Fig. \ref{fig-sonine} and the discussion below. Therefore, similarly as in our previous studies, \cite{SpahnetalEPL:2004,constt1,PNAS,nature}, we approximate the velocity distribution function  for each individual species by the corresponding Maxwell distribution,
\begin{equation}
\label{eq:fk}
  f_k \left( \vec V_k \right) = \frac{n_k}{\left( 2 \pi \theta_k \right)^{3/2}} e^{\frac{\left( \vec V_k - \vec u_k \right)^2}{2 \theta_k}},
\end{equation}
where $n_k (x,t) = \int f_k \left( \vec V_k, \vec x, t \right) d \vec V_k$ is the number density of clusters of size $k$,  $\vec u_k = n_k^{-1} \int \vec V_k f_k d \vec V_k$ is the flux velocity of such clusters, and  $\theta_k = T_k/m_k=\tfrac13 n_k^{-1} \int ( \vec V_k - \vec u_k )^2 f_k d \vec V_k$ is their reduced ``granular'' temperature. Note that the reduced   temperatures, $\theta_k$, describe the speed variance for each partial distribution, $f_k$ \cite{granmix2,bodrova2014}; the average temperature of the whole system is  defined as $T = \sum\limits_k n_k T_k / \sum\limits_k n_k$.
\begin{widetext}
\section{Derivation detail of the hydrodynamic Smoluchowski-Euler equations}
We start from the simplified case when all collisions are aggregative which corresponds to the case of a very large adhesive barrier, $W_{ij}/k_BT \to \infty$ for all $k$. In this case, one can neglect the restitution integrals, $I_{ij}^{{\rm res},1}=I_{ij}^{{\rm res},2}=0$. Integrating equations \eqref{eq:boltz} over the  velocity $\vec V_k$, with the use of the distribution \eqref{eq:fk} we arrive at the Smoluchowski equations for space-inhomogeneous system,

\begin{equation}\label{eq:n2}
  \frac{\partial}{\partial t} n_k + \vec \nabla \left( n_k \vec u_k \right) = \frac{1}{2} \sum\limits_{i+j=k} C_{ij} n_i n_j - \sum\limits_{j = 1}^{\infty} C_{kj} n_k n_j,  \qquad \qquad   \qquad \qquad  \quad  \qquad \qquad   \qquad \qquad \quad k = \overline{1, \infty}.
\end{equation}
Similarly, multiplying Eqs. \eqref{eq:boltz} with the velocity $\vec V_k$, and then with the square of the local velocity, $\tfrac13\vec v_k^{\,2} = \tfrac13 \left| \vec V_k- \vec u_k \right|^2$, and integrating over $\vec V_k$, we obtain,

\begin{align}
  & \frac{\partial}{\partial t} \left( n_k \vec u_k \right) + n_k \vec u_k \cdot \vec \nabla  \vec u_k + \vec u_k \vec \nabla \cdot \left( n_k \vec u_k \right) + \vec \nabla \left( n_k \theta_k \right) -  \frac{\vec F_k }{m_k} n_k
  = \frac{1}{2} \sum\limits_{i+j=k} \vec P_{ij} n_i n_j - \sum\limits_{j = 1}^{\infty} \vec R_{kj} n_k n_j, \quad k = \overline{1, \infty}, \label{eq:u2} \\
  &\frac{\partial}{\partial t} \left( n_k \theta_k \right) +  \vec \nabla \cdot \left( n_k \theta_k \vec u_k \right) + \frac{2}{3} n_k \theta_k \vec \nabla \cdot \vec u_k
  = \frac{1}{2} \sum\limits_{i+j=k} B_{ij} n_i n_j - \sum\limits_{j = 1}^{\infty} D_{kj} n_k n_j,   \qquad \qquad  \qquad   \qquad  \quad k = \overline{1, \infty}. \label{eq:t2}
\end{align}

We can call the system \eqref{eq:n2}-\eqref{eq:t2} the Smoluchowski-Euler equations.  Indeed, when aggregation and energy loss are lacking, the right-hand sides (r.h.s)  disappear, and they convert into Euler equations for the multicomponent molecular fluid. For granular gases, one needs to keep $I_{ij}^{{\rm res},1/2}$ which  results in an additional dissipation term in Eq. \eqref{eq:t2}. In the lack of currents, Eq. \eqref{eq:n2} converts into Smoluchowski equations, if we neglect variation of species temperature; otherwise it converts into temperature-dependent Smoluchowski equations \cite{OsBrilJPA2022,OsBrilPRE2022,gen-smol}.

Surprisingly, a set of  new kinetic coefficients appear -- the vectorial $\vec P_{ij}$ and $\vec R_{ij}$ and scalar -- $B_{ij}$ and $D_{ij}$ (see the derivation below). They reflect the aggregation kinetics in the presence of currents and hence may be dubbed as {\em ``flux-reaction''} and {\em ``energy-reaction''} rates.
Since all the kernels $C_{ij}$, $\vec P_{ij}$, $\vec R_{ij}$, $B_{ij}$ and $D_{ij}$ in \eqref{eq:u2}-\eqref{eq:t2} can  be found analytically, we have a closed set for $n_k$, $\vec u_k$ and $T_k$, that can be numerically solved, similarly as classical Smoluchowski equations.

If the flow speeds and temperatures are the same for all sizes, and all collisions lead to aggregation, we get the standard ballistic kernel \cite{equaltemporig,constt1}:
\[
  C_{ij} = 2 \sqrt{2 \pi} \sigma_{ij}^2 \sqrt{\frac{T}{i} + \frac{T}{j}}.
\]
If not all collisions are aggregative and not all temperatures are equal (although all flow speeds are still the same), the above relation  generalizes to,
\[
  C_{ij} = 2 \sqrt{2 \pi} \sigma_{ij}^2 \sqrt{\frac{T_i}{i} + \frac{T_j}{j}} p_{ij} \left( T_i, T_j \right),
\]
where $p_{ij}(T_i, T_j)$ is the aggregation probability, which can be found explicitly \cite{nature}. This probability is also sometimes called coagulation efficiency \cite{Seinfeld}.

To obtain the left-hand-side (l.h.s.) of Eq. \eqref{eq:n2}, we make the following transformations (below we assume that $\vec F_k$ does not depend on $\vec V_k$):
$$
\int \left( \frac{\partial}{\partial t} + \vec V_k \cdot \frac{\partial}{\partial \vec r} + \frac{\vec F_k}{m_k} \cdot \frac{\partial}{\partial \vec V_k} \right) f_k d
 \vec V_k=\frac{\partial}{\partial t} \int f_kd\vec V_k + \frac{\partial}{\partial \vec r} \cdot \int \vec V_k f_kd\vec V_k + \frac{\vec F_k}{m_k} \int \frac{\partial}{\partial \vec V_k} f_kd\vec V_k = \frac{\partial n_k }{\partial t} + \vec \nabla \cdot (n_k \vec u_k) +0,
$$
where we use the definition of $n_k$ and $\vec u_k$. Similarly, we obtain the $\beta$-component ($\alpha, \beta =x,y,z$) of the l.h.s. of Eq. \eqref{eq:u2}, using the summation convention and $\vec v_k= \vec V_k -\vec u_k$,
\begin{eqnarray}
\label{eq:Vkfk}
&& \int  V_{k,\beta} \left( \frac{\partial}{\partial t} +  V_{k,\alpha}  \frac{\partial}{\partial r_{\alpha}} + \frac{F_{k,\alpha}}{m_k} \cdot \frac{\partial}{\partial  V_{k,\alpha}} \right) f_k d
 \vec V_k \nonumber \\
 &&~~~= \frac{\partial}{\partial t} \int V_{k,\beta}  f_kd\vec V_k + \frac{\partial}{\partial  r_{\alpha}} \int  V_{k,\alpha} V_{k,\beta} f_kd\vec V_k + \frac{F_{k,\alpha}}{m_k} \int  V_{k,\beta } \frac{\partial}{\partial V_{k,\alpha}} f_kd\vec V_k  \nonumber \\
&& ~~~= \frac{\partial  }{\partial t} (n_k u_{k,\beta}) +\frac{\partial}{\partial r_{\alpha}}  \int (v_{k,\alpha}+u_{k,\alpha})(v_{k,\beta}+u_{k,\beta}) f_kd\vec V_k - \frac{F_{k,\alpha}}{m_k} \delta_{\alpha,\beta} \int f_kd\vec V_k
\nonumber \\
&& ~~~= \frac{\partial  }{\partial t} (n_k u_{k,\beta}) +\frac{\partial}{\partial r_{\alpha}} \delta_{\alpha,\beta} \int v_{k,\beta} v_{k,\alpha} f_kd\vec V_k   +\frac{\partial}{\partial r_{\alpha}} (u_{k,\beta}u_{k,\alpha} n_k )  - \frac{F_{k,\beta}}{m_k} n_k  \nonumber \\
&& ~~~= \frac{\partial  }{\partial t} (n_k u_{k,\beta}) +\frac{\partial}{\partial r_{\beta}} n_k \theta_k + u_{k,\beta} \vec \nabla \cdot (n_k \vec u_k) + n_k (\vec u_k \cdot \vec \nabla) u_{k,\beta}-\frac{F_{k,\beta}}{m_k} n_k .
 \end{eqnarray}
Here we also used $\int (v_{k,\alpha})^n f_k d \vec V_k=0$ for odd $n$. For the l.h.s. of Eq. \eqref{eq:t2} we find:

\begin{eqnarray}
\label{eq:fkVk2}
&& \int  \frac{v_{k}^2}{3} \left( \frac{\partial}{\partial t} +  V_{k,\alpha}  \frac{\partial}{\partial r_{\alpha}} + \frac{F_{k,\alpha}}{m_k} \cdot \frac{\partial}{\partial  V_{k,\alpha}} \right) f_k d
 \vec V_k \nonumber \\
 &&~~~= \int \frac13 (\vec{V}_k-\vec{u}_k)^2 \left( \frac{\partial}{\partial t} +  V_{k,\alpha}  \frac{\partial}{\partial r_{\alpha}} \right) f_k d \vec V_k  + \frac{F_{k,\alpha}}{m_k} \int \frac{v_k^2}{3} \frac{\partial}{\partial V_{k,\alpha}} f_k d\vec V_k  \nonumber \\
&& ~~~= \frac13 \frac{\partial  }{\partial t} \int V_k^2 f_k d\vec V_k  - \frac23 \vec{u}_k \cdot \frac{\partial  }{\partial t} \int \vec{V}_k  f_k d\vec V_k  +
\frac13 u_k^2 \frac{\partial n_k }{\partial t} + \frac13 \frac{\partial}{\partial r_{\alpha}} \int V_k^2 V_{k,\alpha}f_k d\vec V_k \nn \\
&& ~~~-\frac23 \vec{u}_k \cdot \frac{\partial}{\partial r_{\alpha}} \int V_{k,\alpha} \vec{V}_k  f_k d\vec V_k +\frac13 u_k^2 \frac{\partial}{\partial r_{\alpha}} n_k u_{k,\alpha}- \frac{F_{k,\alpha}}{m_k}  \int  f_k\left(\frac{\partial}{\partial V_{k,\alpha}} \frac{v_k^2}{3} \right)  d\vec V_k
\nonumber \\
&& ~~~= \frac{\partial  }{\partial t} (n_k \theta_k ) + \vec \nabla \cdot (n_k \theta_k \vec u_k) +\frac23 n_k\theta_k \vec{\nabla} \cdot \vec{u}_k.
 \end{eqnarray}
In the case when the external force depends on the velocity $\vec V_k$, that is,  $\vec F_k=\vec h_k (\vec V_k)$, the last term in Eq. \eqref{eq:dfdt} should be written as $\partial/\partial \vec V_k  \cdot (f_k\vec F_k/m_k)$. Then the last term in the r.h.s. of Eq. \eqref{eq:Vkfk} takes the form $-\int f_k \vec h_k \cdot d \vec V_k$, or $3\pi \eta \sigma_k n_k \vec u_k /m_k$, when $\vec F_k$ obeys the Stokes law. Similarly, in the r.h.s. of  Eq. \eqref{eq:fkVk2} appears an additional term, $6\pi \eta \sigma_k n_k \theta_k/m_k$, if $\vec F_k$ follows the Stokes law.

Turn now to the derivation of the r.h.s. of Eqs. \eqref{eq:n2}-\eqref{eq:t2} which implies the derivation of the respective kinetic coefficients.

\section{Derivation of the kinetic coefficient for all-merging  collisions}\label{sec:deriv}
\subsection{3D case}
Let us derive the kernels (i.e. the transport-reactive coefficients) for the case, when all collisions are merging.  
Here we present the detailed, step-by-step derivation for the kinetic coefficients $C_{ij}$; all other kinetic coefficients $\vec P_{ij}$, $\vec R_{ij}$, $B_{ij}$ and $D_{ij}$ may be derived analogously.

First, we notice that the following relation holds true,
\begin{equation}\label{eq:cijdef}
  C_{ij} n_i n_j = \int I_{ij}^{{\rm agg}, 2} \, d \vec V_i,
\end{equation}
that is, we need to integrate over all speeds to find the total number of collisions which in our case is the same as the total number of aggregating collisions. As before, $\vec v_i = \vec V_i - \vec u_i$ will be the ``local'' speed of size-$i$ particles -- the speed in the system of coordinates moving with their flux  velocity $\vec u_i$. Hereinafter, we will use $m_i = i$. To simplify the  computations we introduce the scaled temperatures $\theta_i = T_i / i$.

We start with the change of variables in $I_{ij}^{{\rm agg}, 2}$ from $\vec V_i$ and $\vec V_j$ to
\[
\begin{aligned}
  \vec w & = \vec v_i - \vec v_j, \\
  \vec u & = \frac{\theta_j \vec v_i + \theta_i \vec v_j}{\theta_i + \theta_j}
\end{aligned}
\]
and denote
\[
  \vec W = \vec V_i - \vec V_j.
\]
Substituting the Maxwell distribution and using the new variables yields,
\begin{equation}
    \label{Cij_1}
  C_{ij} = \frac{\sigma_{ij}^2}{\left( 2 \pi \right)^3 \theta_i^{3/2} \theta_j^{3/2}} \iiint d \vec w \, d \vec u \, d \vec e \, \Theta \left( - \vec W \cdot \vec e \right) \left| \vec W \cdot \vec e \right| e^{- \frac{u^2}{2} \left( \theta_i^{-1} + \theta_j^{-1} \right) - \frac{w^2}{2 \left( \theta_i + \theta_j \right)}}.
\end{equation}
The above integral is Gaussian with respect to $\vec u$ and hence may be easily calculated, with the result:
\begin{equation}
    \label{Cij_2}
  C_{ij} = \frac{\sigma_{ij}^2}{\left( 2 \pi \right)^{3/2} \left( \theta_i + \theta_j \right)^{3/2}} \iint d \vec w \, d \vec e \, \Theta \left( - \vec W \cdot \vec e \right) \left| \vec W \cdot \vec e \right| e^{- \frac{w^2}{2 \left( \theta_i + \theta_j \right)}}.
\end{equation}
Integration over the unit vector $\vec e$ (actually only over the semi-sphere) may be also easily performed  (see e.g. Ref. \cite{BrilliantovPoeschelOUP} where such integrals are evaluated). With the obvious notation for vectors moduli, $W = \left| \vec W \right|$, we arrive at:
\[
  C_{ij} = \frac{\pi \sigma_{ij}^2}{\left( 2 \pi \right)^{3/2} \left( \theta_i + \theta_j \right)^{3/2}} \int d \vec w \, W e^{- \frac{w^2}{2 \left( \theta_i + \theta_j \right)}}.
\]
Next, we make another change of variables $\vec v = \vec W / \sqrt{2 \left( \theta_i + \theta_j \right)}$ and  denote
\[
  \vec c = \sqrt{\frac{2}{\theta_i + \theta_j}} \left( \vec u_i - \vec u_j \right).
\]
Then
\begin{equation}
    \label{Cij_3}
  C_{ij} = \sqrt{\frac{2}{\pi}} \sigma_{ij}^2 \sqrt{\theta_i + \theta_j} \int d \vec v \, v e^{- \left| \vec v - \frac{\vec c}{2} \right|^2}.
\end{equation}
Changing to the cylindrical coordinates $\left( h, r, \phi \right)$, where $h$ is the coordinate for the axis directed along $\vec c$, and integrating over the angle $\varphi$ we obtain:
\[
  C_{ij} = 2 \sqrt{2 \pi} \sigma_{ij}^2 \sqrt{\theta_i + \theta_j} \int_{-\infty}^{\infty} d h \int_0^{\infty} d r \, r\,\sqrt{r^2 + h^2} e^{- r^2 - \left| h - \frac{c}{2} \right|^2}.
\]
Now we change the variable in the second integral as $R = r^2 + h^2$ and integrate by parts
\begin{equation}
    \label{Cij_4}
\begin{aligned}
  C_{ij} & = \sqrt{2 \pi} \sigma_{ij}^2 \sqrt{\theta_i + \theta_j} \int_{-\infty}^{\infty} d h \int_{h^2}^{\infty} d R \, \sqrt{R} e^{h^2 - R - \left| h - \frac{c}{2} \right|^2} \\
  & = 2 \sqrt{2 \pi} \sigma_{ij}^2 \sqrt{\theta_i + \theta_j} \int_{-\infty}^{\infty} \frac{h \left| h \right|}{c} e^{-h^2 + ch - c^2/4} \, dh.
\end{aligned}
\end{equation}
The above integral may be written as a sum of two parts,  corresponding to the different signs of $h$, as:
\[
  C_{ij} = 2 \sqrt{2 \pi} \sigma_{ij}^2 \sqrt{\theta_i + \theta_j} \left( \int_{0}^{\infty} \frac{h^2}{c} e^{-h^2 + ch - c^2/4} \, dh - \int_{0}^{\infty} \frac{h^2}{c} e^{-h^2 - ch - c^2/4} \, dh \right).
\]
After the change of variables $z = h - c/2$, both integrals will turn into a combination of incomplete Gamma functions of an integer argument, yielding  finally the result
\begin{equation}\label{eq:cagg0}
\begin{aligned}
  & C_{ij} = \sqrt{2 \pi} \sigma_{ij}^2 \sqrt{\theta_i + \theta_j} \left[e^{-c^2/4} + \frac{\sqrt{\pi} (c^2+2)}{2c}{\rm erf}(\tfrac{c}{2}) \right] ,  \\
  & c = \sqrt{\frac{2}{\theta_i + \theta_j}} \left| \vec u_i - \vec u_j \right|.
\end{aligned}
\end{equation}
Though the integral in Eq. \eqref{Cij_4} is undefined for $\vec u_i = \vec u_j$, i.e., for $c = 0$, the corresponding value can still be found in the limit $c \to +0$ and coincides with the flux-less case.
Using the same steps as for the derivation of $C_{ij}$, we find all other kinetic coefficients:
\[
\begin{aligned}
  \vec P_{ij} & = C_{ij} \vec \mu + 2 \sqrt{2 \pi} \sigma_{ij}^2 q \vec c \left( e^{-c^2/4} \left( 1/c^2 + 6 \right) + \frac{\sqrt{\pi}}{2c} \left( 1 - 2/c^2 \right) {\rm erf} \left( c/2 \right) \right), \\
  \vec R_{ij} & = C_{ij} \vec u_i + 2 \sqrt{2 \pi} \sigma_{ij}^2 \theta_i \vec c \left( e^{-c^2/4} \left( 1/c^2 + 6 \right) + \frac{\sqrt{\pi}}{2c} \left( 1 - 2/c^2 \right) {\rm erf} \left( c/2 \right) \right), \\
  B_{ij} & = C_{ij} \left( \left| \vec \mu - \vec u_{i+j} \right|^2 / 3 + \frac{\theta_i \theta_j}{\theta_i + \theta_j} - \frac{q}{3 \left( \theta_i + \theta_j \right)} \left( \vec u_i - \vec u_j \right) \cdot \left( \vec \mu - \vec u_{i+j} \right) \right) \\
  & + \frac23 \sqrt{2 \pi} \sigma_{ij}^2 \sqrt{\theta_i + \theta_j} q I_B \left| \vec \mu - \vec u_{i+j} \right| / \left| \vec u_i - \vec u_j \right| \\
  & + \sqrt{2 \pi} \sigma_{ij}^2 q^2 \left( e^{-c^2/4} + \frac{\sqrt{\pi}}{6c} \left( 10 + 3c^2 \right) {\rm erf} \left( c/2 \right) \right) / \sqrt{\theta_i + \theta_j}, \\
  D_{ij} & = C_{ij} \frac{\theta_i \theta_j}{\theta_i + \theta_j} + \sqrt{2 \pi} \sigma_{ij}^2 \frac{\theta_i^2}{\sqrt{\theta_i + \theta_j}} \left( e^{-c^2/4} + \frac{\sqrt{\pi}}{6c} \left( 10 + 3c^2 \right) {\rm erf} \left( c/2 \right) \right), \\
  I_B & = \frac{\sqrt{\pi}}{8} \left( c^3 + 4c - 4/c \right) {\rm erf} \left( c/2 \right) + \frac12 e^{-c^2/4} + \frac{13}{4} c^2 e^{-c^2/4}, \\
  \vec \mu & = \frac{i \vec u_i + j \vec u_j}{i + j}, \\
  q & = \frac{T_i - T_j}{i + j}.
\end{aligned}
\]
If the particles are spherical and monomers have a unit diameter, then ${\sigma _{ij}} = \tfrac12(\sigma_i + \sigma_j)= \tfrac12(i^{1/3} + j^{1/3})$.

\subsection{Quasi-1D case}

Previously, we assumed that the speed variance is spherically symmetric $\theta_{k,x} = \theta_{k,y} = \theta_{k,z}$. There exists, however, an important for applications case of quasi-one dimensional motion,  when one of the components dominates, e.g. $\theta_{k,z} = \int v_{k,z}^2 f_k \left( \vec v_k, \vec x, t \right) d \vec v_k \gg \max \left( \theta_{k,x}, \theta_{k,y} \right)$.  As an example of such systems, one can mention aggregating particles freely falling in the air. Generally,  the particles experience three forces -- the force of gravity (along with the buoyancy), the force of viscous resistance and the stochastic force, due to random collisions of the particles with the molecules of the gas. If the particles are massive enough,  one can neglect the stochastic force,  as the change of particles' momentum  in the collisions with the gas molecules is negligible. In this case, the lateral motion of particles is quickly damped by the air viscosity, while the gravity force supports the vertical motion with a high but constant velocity. This makes the motion of the system quasi-one dimensional. The dispersion of the cluster velocities, quantified by the partial temperatures $T_k$, emerges due to collisions between the aggregates.

The derivation of the kinetic coefficients is similar in this case, although the resulting kernels are different. We choose the coordinate system with the vertically directed axis $OZ$. In this case, the horizontal velocity variance is quickly dumped by the air resistance. The same happens with the corresponding flux velocities, so that $u_{i,x} = u_{i,y}=0$ for all $i$. With the notations $V_k = V_{k,z}$, $u_k = u_{k,z}$ and $\theta_k = \theta_{k,z}$, we obtain,
\[
  f_k \left( \vec V_k \right) \approx \frac{n_k}{\left( 2 \pi \theta_k \right)^{1/2}} e^{\frac{\left( V_k - \vec u_k \right)^2}{2 \theta_k}} \delta \left( V_{k,x} \right) \delta \left( V_{k,y} \right).
\]
We use similar variables as before, although these are now one-dimensional:
\[
\begin{aligned}
  w & = v_i - v_j, \\
  u & = \frac{\theta_j v_i + \theta_i v_j}{\theta_i + \theta_j}, \\
  W & = V_i - V_j.
\end{aligned}
\]
Substituting the distribution functions into the equation for $C_{ij}$ leads to
\[
\begin{aligned}
  C^{(1D)}_{ij} & = \frac{\sigma_{ij}^2}{2 \pi \theta_i^{1/2} \theta_j^{1/2}} \iiint d w \, d u \, d \vec e \, \theta \left( - \vec W \cdot \vec e \right) \left| \vec W \cdot \vec e \right| e^{- \frac{u^2}{2} \left( \theta_i^{-1} + \theta_j^{-1} \right) - \frac{w^2}{2 \left( \theta_i + \theta_j \right)}} \\
  & = \frac{\sigma_{ij}^2}{ \left( 2 \pi \right)^{1/2} \left( \theta_i + \theta_j \right)^{1/2}} \iiint d w \, d \vec e \, \theta \left( - \vec W \cdot \vec e \right) \left| \vec W \cdot \vec e \right| e^{ - \frac{w^2}{2 \left( \theta_i + \theta_j \right)}} \\
  & = \frac{\pi \sigma_{ij}^2}{ \left( 2 \pi \right)^{1/2} \left( \theta_i + \theta_j \right)^{1/2}} \iiint d w \, \left| W \right| e^{ - \frac{w^2}{2 \left( \theta_i + \theta_j \right)}}.
\end{aligned}
\]
Integrating in the same way as before, we get the following kernel:
\begin{equation}\label{eq:cagg2}
\begin{aligned}
  & C^{(1D)}_{ij} = \sqrt{2 \pi} \sigma_{ij}^2 \sqrt{\theta_i + \theta_j} e^{-c^2/4} + \pi \sigma_{ij}^2 \left| u_i - u_j \right| {\rm erf} \left( c/2 \right), \\
  & c = \sqrt{\frac{2}{\theta_i + \theta_j}} \left| \vec u_i - \vec u_j \right|.
\end{aligned}
\end{equation}
Other kernels for the quasi-one dimensional case read:
\[
\begin{aligned}
  P^{(1D)}_{ij} & = C_{ij} \mu + \pi \sigma_{ij}^2 q {\rm erf} \left( c/2 \right), \\
  R^{(1D)}_{ij} & =  C_{ij} u_i + \pi \sigma_{ij}^2 \theta_i {\rm erf} \left( c/2 \right), \\
  B^{(1D)}_{ij} & = C_{ij} \left( \left| \vec \mu - \vec u_{i+j} \right|^2 + \frac{\theta_i \theta_j + q^2}{\theta_i + \theta_j} \right) \\
  & + 2 \pi \left( \mu - u_{i+j} \right) \sigma_{ij}^2 q {\rm erf} \left( c/2 \right) + \sqrt{2 \pi} \sigma_{ij}^2 q^2 e^{-c^2/4} / \sqrt{\theta_i + \theta_j}, \\
  D^{(1D)}_{ij} & = C_{ij} \theta_i + \sqrt{2 \pi} \sigma_{ij}^2 \theta_i^2 e^{-c^2/4} / \sqrt{\theta_i + \theta_j}, \\
  \mu & = \frac{i u_i + j u_j}{i + j}, \\
  q & = \frac{T_i - T_j}{i + j}.
\end{aligned}
\]
The terms that stem from the source term ${\cal J}$ in the equations for the quasi-one dimensional case on the r.h.s. of these equations read,
$$
\int J \delta(z) [\delta_{k,1} \delta (V_1-V_{\rm eq, 1})+\delta_{k,2} \delta (V_1-V_{\rm eq, 2})] f_k d\vec V_k = J \delta(z) [\delta_{k,1} +\delta_{k,2} ],
$$
$$
\int J \delta(z) [\delta_{k,1} \delta (V_1-V_{\rm eq, 1})+\delta_{k,2} \delta (V_1-V_{\rm eq, 2})] V_k f_k d\vec V_k = J \delta(z) [\delta_{k,1} + 2^{2/3} \delta_{k,2} ],
$$
and
$$
\int J \delta(z) [\delta_{k,1} \delta (V_1-V_{\rm eq, 1})+\delta_{k,2} \delta (V_1-V_{\rm eq, 2})] \tfrac13 v_k^2 f_k d\vec V_k =  0,
$$
since we assume that the particles of the source have the same steady velocity, $V_k=V_{\rm eq, k}= u_k$, which implies $v_k=0$.

We can also write the Smoluchowski-Euler equations in spherical coordinates, where we take into account that the radial component of the  gradient of a function $H$ reads $\vec e_r (\partial H/\partial r)$, while divergence of a vector $\vec A= (A_r, 0,0)$ is $r^{-2} \partial (r^2 A_r)/\partial r $.

Let us compare our results with some previous models. We have already seen that our kernel $C_{ij}$ is a direct generalization of the classical ballistic kernel. However, we should also look at the previous attempts to add flow speeds into the equations. In the simplest case, the universal flow speed and temperature can be introduced phenomenologically, like in Ref. \cite{proto}. The authors write the coagulation kernel as
\begin{equation}\label{eq:planet}
  C_{ij} = 2 \sqrt{2 \pi} \sigma_{ij}^2 \sqrt{\frac{T}{i} + \frac{T}{j}} + \pi \sigma_{ij}^2 \left| \vec u_i - \vec u_j \right|,
\end{equation}

just by adding the contributions of flow speed and thermal speed. Naturally, this gives the right prediction, when either flow speed or thermal speed dominates, which is also true for our results, Eq. \eqref{eq:cagg0}. However, the simple model \eqref{eq:planet} cannot be used when flow speed differences and thermal speed differences are of the same order. Also, it cannot be used for granular gases or mixtures of granular and molecular gases, when the temperatures of different species are not the same, or temperatures of granular particles differ from that of molecular gas. This is commonly the case for ballistic aggregation: when there are only a few particles of granular gas, their rare collisions do not affect their temperature much; instead, they experience Brownian motion in molecular gas. In the case of ballistic motion, granular temperature is much smaller than that of molecular gas, even in non-aggregating granular gases due to energy loss in their inelastic collisions \cite{bodrovamol}.

Therefore, new equations reported in the main part of the article solve several problems. They are applicable independently of the ratio of the thermal and flow velocities and can account for the temperature difference of different species, which is a common case in driven granular gases \cite{bodrova}. They account for the difference between granular temperatures and the temperature of the medium, and they can be used to predict the temperature and flow speed evolution in granular gases, as they change during the aggregation. The correct description of the evolution of the entangled number densities, flow fields and temperatures of different species provides a more accurate prediction for the behavior of the system, as compared to that based on the equations with phenomenological rate coefficients.

\section{Derivation of the kinetic coefficient for arbitrary potential barrier}\label{sec:deriv_general}
The derivation of the kinetic coefficients for the general case of an arbitrary value of the potential barrier $W_{ij}$ follows the same lines as for the case of all-merging collisions. The only difference is the appearance in the integrands of the additional factor, $\Theta \left(W_{ij} - \tfrac12\varepsilon^2 \mu_{ij} (\vec{V}_i- \vec{v}_j)^2 \right)$, which selects merging collisions from all possible (including bouncing) collisions. In particular, this factor appears in the derivation of the kinetic coefficients $C_{ij}$, so that Eq. \eqref{Cij_1} now reads,
\begin{equation}
    \label{Cij_1gen}
  C_{ij} = \frac{\sigma_{ij}^2}{\left( 2 \pi \right)^3 \theta_i^{3/2} \theta_j^{3/2}} \iiint d \vec w \, d \vec u \, d \vec e \, \Theta \left( - \vec W \cdot \vec e \right)
  \Theta \left( W_{ij} - \tfrac12 \varepsilon^2 \mu_{ij} \vec W^2 \right)
  \left| \vec W \cdot \vec e\,  \right| e^{- \frac{u^2}{2} \left( \theta_i^{-1} + \theta_j^{-1} \right) - \frac{w^2}{2 \left( \theta_i + \theta_j \right)}},
\end{equation}
which transforms, after the integration over $\vec u$,  into,
\begin{equation}
    \label{Cij_2gen}
  C_{ij} = \frac{\sigma_{ij}^2}{\left( 2 \pi \right)^{3/2} \left( \theta_i + \theta_j \right)^{3/2}} \iint d \vec w \, d \vec e \, \Theta \left( - \vec W \cdot \vec e \right) \Theta \left( W_{ij} - \tfrac12 \varepsilon^2 \mu_{ij} \vec W^2 \right)
  \left| \vec W \cdot \vec e \, \right| e^{- \frac{w^2}{2 \left( \theta_i + \theta_j \right)}}.
\end{equation}
Making the same transformations which lead to Eq. \eqref{Cij_3}, we obtain, instead of Eq. \eqref{Cij_3}:
\begin{equation}
    \label{Cij_3gen}
  C_{ij} = \sqrt{\frac{2}{\pi}} \sigma_{ij}^2 \sqrt{\theta_i + \theta_j} \int d \vec v \, \Theta(q_{ij}-v^2) \, v \,e^{- \left| \vec v - \frac{\vec c}{2} \right|^2},
\end{equation}
where we abbreviate, $q_{ij}= W_{ij}/(\varepsilon^2 \mu_{ij} (\theta_i+\theta_j))$ and $c=\sqrt{2/(\theta_i+\theta_j)}\left| \vec u_i - \vec u_j\right|$. With the same steps as before, the above relation may be further transformed into the following counterpart of Eq. \eqref{Cij_4}:
\begin{equation}
    \label{Cij_4gen}
  C_{ij} =2 \sqrt{2 \pi} \sigma_{ij}^2 \sqrt{\theta_i + \theta_j} \int_{-\sqrt{q_{ij}}}^{\sqrt{q_{ij}}} \frac{h \left| h \right|}{c} e^{-h^2 + ch - c^2/4} \, dh.
\end{equation}
The above integral may be evaluated and expressed in terms of incomplete gamma functions. The final result for $C_{ij}$ reads:
\begin{equation}
    \label{Cij_5gen}
{C_{ij}} = 2\sqrt {2\pi } \sigma _{ij}^2\sqrt {{\theta _i} + {\theta _j}} \, I_F \left( {q_{ij}}, \vec c  \, \right)= C_{ij}^{(0)} I_F \left( {q_{ij}}, \vec c  \, \right),
\end{equation}
where $C_{ij}^{(0)}$ corresponds to the case of all-merging collisions in homogeneous systems without fluxes. $I_F \left( {q_{ij}}, \vec c  \, \right)$, with $\vec c=\sqrt{2/(\theta_i+\theta_j)}\left( \vec u_i - \vec u_j\right)$,  and similar integrals, will be presented below. Note, that for $q \to \infty$, that is, for the infinitely large barrier, when all collisions are merging,  $I_F(\infty, \vec c) = \left[2c \, e^{-c^2/4} + \sqrt{\pi} (c^2+2) {\rm erf}(c/2)\right]/4c$,  and the previous result, Eq. \eqref{eq:cagg0}, is recovered.

Similarly, in the derivation of the coefficients $\vec{P}_{ij}$, $\vec{R}_{ij}$, $B_{ij}$ and $D_{ij}$ for arbitrary $W_{ij}$,  one can make the same steps as in the calculation  of $C_{ij}$, for the case of $W_{ij} \to \infty$,  up to the very last step. At the last step -- the respective integration is to be performed in the limits $(-\sqrt{q_{ij}}, \sqrt{q_{ij}})$, instead of the limits $(-\infty, \infty)$ for the case of $W_{ij} \to \infty$. The computations associated with the other parts of the collision integral, $I_{ij}^{\rm res, 1/2}$ are very similar to that, explained  for $I_{ij}^{\rm agg, 2}$, and  straightforward. Here we present the final results for the vectorial coefficients:
\begin{equation}\label{Cij_6gen}
\begin{gathered}
  {\vec P_{ij}} = C_{ij}^{(0)}  \left[ \frac{i \vec u_i + j \vec u_j}{i + j} I_F \left( {q_{ij}}, \vec c \right) + \left( \vec u_i - \vec u_j \right) \left( {\frac{{i{\theta _i} - j{\theta _j}}}{{i + j}}} \right) \,
  \left( \frac{1}{|\vec u_i - \vec u_j|^2} I_H \left( q_{ij}, \vec c \,\right)  - \frac{1}{\theta_i + \theta_j} I_F \left( q_{ij}, \vec c \,\right) \right)  \right], \hfill \\
  {\vec R_{ij}} = C_{ij}^{(0)} \left[ \vec u_i I_F \left( {q_{ij}}\vec c\,  \right)
+ \left( \vec u_i - \vec u_j \right) \left( \frac{\theta_i}{|\vec u_i - \vec u_j|^2} I_H \left( {q_{ij}}, \vec c \, \right) - \frac{\theta_i}{\theta_i + \theta_j} I_F \left( {q_{ij}}, \vec c\, \right) \right) \right. \hfill \\
  \left.  ~~~~+ \frac{1}{2} \left( \vec u_i - \vec u_j \right) \frac{j}{{i + j}} \left( \theta_i + \theta_j \right) \left( {1 + \varepsilon } \right) \frac{1}{|\vec u_i - \vec u_j|^2} \overline{I_H} \left( {q_{ij}}, \vec c\, \right) \right], \hfill \\
\end{gathered}
\end{equation}
and for the scalar coefficients:
\begin{equation}\label{eq-eulerkernels}
\begin{gathered}
  {B_{ij}} = \frac{C_{ij}^{(0)}}{(\theta _i + \theta _j) } \left[ \theta _i\theta _j I_F \left( q_{ij}, \vec c\,\right) +  \left( \frac{\left( \theta_i + \theta_j \right)^2}{6} \left|\Delta \vec u_{ijk} \right|^2 - \frac{i \theta_i - j \theta_j}{6(i+j)} \left( \theta_i + \theta_j \right) \left( \vec c \cdot \Delta \vec u_{ijk} \right) \right)
  I_F \left( q_{ij}, \vec c \,\right) \right. \hfill \\
   ~~~~~~ + \left. \frac{\left( \theta_i + \theta_j \right)\left(i\theta _i - j\theta _j \right) \left( \vec c \cdot \Delta \vec u_{ijk} \right)}{3(i + j) \left| \vec c \right|^2}  I_{H} \left( q_{ij}, \vec c \,\right)  + \frac{4}{3} \left( \frac{i\theta _i - j\theta _j}{i + j} \right)^2 I_{GH2} \left( q_{ij}, \vec c \,  \right) \right], \hfill \\
  {D_{ij}} = \frac{C_{ij}^{(0)}}{(\theta _i + \theta _j) } \left[\theta _i\theta _jI_F \left( q_{ij}, \vec c \, \right)  + \frac{4}{3}\theta _i^2 I_{GH2} \left( q_{ij}, \vec c \, \right) + \frac{4}{3}\frac{j}{{i + j}} \left(
  \theta_i + \theta_j \right)\left( {1 + \varepsilon } \right) \left( \theta _i \overline{I_{GH}} \left( q_{ij}, \vec c \, \right)  \right.   \right. \hfill \\
    \left.  \left. ~~~~~ - \frac{1}{2}\left( {1 + \varepsilon } \right)\frac{j}{{i + j}}\left( {{\theta _i} + {\theta _j}} \right) \overline{I_G} \left( q_{ij}, \vec c \,\right) \right) \right], \hfill \\
\end{gathered}
\end{equation}
where
\begin{equation}\label{eq:q}
\begin{gathered}
     {q_{ij}} = \frac{W_{ij}}{\varepsilon^2 \, \left( \theta _i + \theta _j \right)(ij)/(i+j)}, \hfill \\
  {\Delta \vec u_{ijk}} = \sqrt{\frac{2}{\theta_i + \theta_j}} \left( \frac{i \vec u_i + j \vec u_j}{i+j} - \vec u_{i+j} \right), \hfill \\
  \end{gathered}
\end{equation}
and, as previously, $C_{ij}^{(0)} =2 \sqrt{2 \pi} \sigma_{i}^2 \sqrt{\theta_i+\theta_j}$ and $\sigma _{ij} =\tfrac12(i^{1/3}+j^{1/3})$, with $m_1=1$ and $\sigma_1=1$. $I_F$, $I_G$, $I_H$, $I_{GH2}$, $\overline{I_G}$, $\overline{I_H}$ and $\overline{I_{GH}}$ denote the integrals of special type, given below:
\[
\begin{aligned}
  {I_F}\left( {Q,\vec c} \right) & = \frac{1}{{2\pi }}\int\limits_{0 < {{\left| {\vec w} \right|}^2} < Q} {\left| {\vec w} \right|{e^{ - {{\left| {\vec w - \frac{{\vec c}}{2}} \right|}^2}}}d\vec w}  \\
   & =  - \frac{{\sqrt \pi  }}{{4\left| {\vec c} \right|}}\left( {{\rm erf}\left( {\sqrt Q  + \frac{{\left| {\vec c} \right|}}{2}} \right) - {\rm erf}\left( {\sqrt Q  - \frac{{\left| {\vec c} \right|}}{2}} \right)} \right) + \frac{{\sqrt \pi  }}{{2\left| {\vec c} \right|}} {\rm erf}\left( {\frac{{\left| {\vec c} \right|}}{2}} \right) \\
   & - \frac{{\sqrt \pi  }}{8}\left| {\vec c} \right|\left( {{\rm erf}\left( {\sqrt Q  + \frac{{\left| {\vec c} \right|}}{2}} \right) - {\rm erf}\left( {\sqrt Q  - \frac{{\left| {\vec c} \right|}}{2}} \right)} \right)  + \frac{{\sqrt \pi  }}{4}\left| {\vec c} \right|{\rm erf}\left( {\frac{{\left| {\vec c} \right|}}{2}} \right) \\
   & + \frac{1}{{2\left| {\vec c} \right|}}\left( {\sqrt Q  + \frac{{\left| {\vec c} \right|}}{2}} \right){e^{ - {{\left( {\sqrt Q  + \frac{{\left| {\vec c} \right|}}{2}} \right)}^2}}} - \frac{1}{{2\left| {\vec c} \right|}}\left( {\sqrt Q  - \frac{{\left| {\vec c} \right|}}{2}} \right){e^{ - {{\left( {\sqrt Q  - \frac{{\left| {\vec c} \right|}}{2}} \right)}^2}}} \\
   & + \frac{1}{2}{e^{ - \frac{{{{\left| {\vec c} \right|}^2}}}{4}}} - \frac{1}{2}{e^{ - {{\left( {\sqrt Q  + \frac{{\left| {\vec c} \right|}}{2}} \right)}^2}}} - \frac{1}{2}{e^{ - {{\left( {\sqrt Q  - \frac{{\left| {\vec c} \right|}}{2}} \right)}^2}}},
\end{aligned}
\]
\[
\begin{aligned}
  \overline{I_F}\left( {Q,\vec c} \right) & = \frac{1}{{2\pi }}\int\limits_{{{\left| {\vec w} \right|}^2} > Q} {\left| {\vec w} \right|{e^{ - {{\left| {\vec w - \frac{{\vec c}}{2}} \right|}^2}}}d\vec w}  \\
   & =  \frac{{\sqrt \pi  }}{{4\left| {\vec c} \right|}}\left( {{\rm erf} \left( {\sqrt Q  + \frac{{\left| {\vec c} \right|}}{2}} \right) - {\rm erf}\left( {\sqrt Q  - \frac{{\left| {\vec c} \right|}}{2}} \right)} \right) + \frac{{\sqrt \pi  }}{8}\left| {\vec c} \right|\left( {{\rm erf}\left( {\sqrt Q  + \frac{{\left| {\vec c} \right|}}{2}} \right) - {\rm erf}\left( {\sqrt Q  - \frac{{\left| {\vec c} \right|}}{2}} \right)} \right) \\
   & - \frac{1}{{2\left| {\vec c} \right|}}\left( {\sqrt Q  + \frac{{\left| {\vec c} \right|}}{2}} \right){e^{ - {{\left( {\sqrt Q  + \frac{{\left| {\vec c} \right|}}{2}} \right)}^2}}} + \frac{1}{{2\left| {\vec c} \right|}}\left( {\sqrt Q  - \frac{{\left| {\vec c} \right|}}{2}} \right){e^{ - {{\left( {\sqrt Q  - \frac{{\left| {\vec c} \right|}}{2}} \right)}^2}}}  + \frac{1}{2}{e^{ - {{\left( {\sqrt Q  + \frac{{\left| {\vec c} \right|}}{2}} \right)}^2}}} + \frac{1}{2}{e^{ - {{\left( {\sqrt Q  - \frac{{\left| {\vec c} \right|}}{2}} \right)}^2}}},
\end{aligned}
\]
\[
\begin{aligned}
  {I_G}\left( {Q,\vec c} \right) & = \frac{1}{{4\pi }}\int\limits_{0 < {{\left| {\vec w} \right|}^2} < Q} {{{\left| {\vec w} \right|}^3}{e^{ - {{\left| {\vec w - \frac{{\vec c}}{2}} \right|}^2}}}d\vec w}  \\
   & =  - \frac{{3\sqrt \pi  }}{{16\left| {\vec c} \right|}}\left( {{\rm erf}\left( {\sqrt Q  + \frac{{\left| {\vec c} \right|}}{2}} \right) - {\rm erf}\left( {\sqrt Q  - \frac{{\left| {\vec c} \right|}}{2}} \right)} \right) + \frac{{3\sqrt \pi  }}{{8\left| {\vec c} \right|}}{\rm erf}\left( {\frac{{\left| {\vec c} \right|}}{2}} \right) \\
   & - \frac{{3\sqrt \pi  }}{{16}}\left| {\vec c} \right|\left( {{\rm erf}\left( {\sqrt Q  + \frac{{\left| {\vec c} \right|}}{2}} \right) - {\rm erf}\left( {\sqrt Q  - \frac{{\left| {\vec c} \right|}}{2}} \right)} \right) + \frac{{3\sqrt \pi  }}{8}\left| {\vec c} \right|{\rm erf} \left( {\frac{{\left| {\vec c} \right|}}{2}} \right) \\
   & - \frac{{\sqrt \pi  }}{{64}}{\left| {\vec c} \right|^3}\left( {{\rm erf}\left( {\sqrt Q  + \frac{{\left| {\vec c} \right|}}{2}} \right) - {\rm erf}\left( {\sqrt Q  - \frac{{\left| {\vec c} \right|}}{2}} \right)} \right) + \frac{{\sqrt \pi  }}{{32}}{\left| {\vec c} \right|^3}{\rm erf}\left( {\frac{{\left| {\vec c} \right|}}{2}} \right) \\
   & + \frac{3}{{8\left| {\vec c} \right|}}\left( {\sqrt Q  + \frac{{\left| {\vec c} \right|}}{2}} \right){e^{ - {{\left( {\sqrt Q  + \frac{{\left| {\vec c} \right|}}{2}} \right)}^2}}} - \frac{3}{{8\left| {\vec c} \right|}}\left( {\sqrt Q  - \frac{{\left| {\vec c} \right|}}{2}} \right){e^{ - {{\left( {\sqrt Q  - \frac{{\left| {\vec c} \right|}}{2}} \right)}^2}}} \\
   & + \frac{1}{{4\left| {\vec c} \right|}}{\left( {\sqrt Q  + \frac{{\left| {\vec c} \right|}}{2}} \right)^3}{e^{ - {{\left( {\sqrt Q  + \frac{{\left| {\vec c} \right|}}{2}} \right)}^2}}} - \frac{1}{{4\left| {\vec c} \right|}}{\left( {\sqrt Q  - \frac{{\left| {\vec c} \right|}}{2}} \right)^3}{e^{ - {{\left( {\sqrt Q  - \frac{{\left| {\vec c} \right|}}{2}} \right)}^2}}} \\
   & + \frac{3}{8}\left| {\vec c} \right|\left( {\sqrt Q  + \frac{{\left| {\vec c} \right|}}{2}} \right){e^{ - {{\left( {\sqrt Q  + \frac{{\left| {\vec c} \right|}}{2}} \right)}^2}}} - \frac{3}{8}\left| {\vec c} \right|\left( {\sqrt Q  - \frac{{\left| {\vec c} \right|}}{2}} \right){e^{ - {{\left( {\sqrt Q  - \frac{{\left| {\vec c} \right|}}{2}} \right)}^2}}} + \frac{5}{8}{e^{ - \frac{{{{\left| {\vec c} \right|}^2}}}{4}}} + \frac{{13}}{{16}}{\left| {\vec c} \right|^2}{e^{ - \frac{{{{\left| {\vec c} \right|}^2}}}{4}}} \\
   & - \frac{1}{2}{\left( {\sqrt Q  + \frac{{\left| {\vec c} \right|}}{2}} \right)^2}{e^{ - {{\left( {\sqrt Q  + \frac{{\left| {\vec c} \right|}}{2}} \right)}^2}}} - \frac{1}{2}{\left( {\sqrt Q  - \frac{{\left| {\vec c} \right|}}{2}} \right)^2}{e^{ - {{\left( {\sqrt Q  - \frac{{\left| {\vec c} \right|}}{2}} \right)}^2}}} \\
   & - \frac{1}{2}{e^{ - {{\left( {\sqrt Q  + \frac{{\left| {\vec c} \right|}}{2}} \right)}^2}}} - \frac{1}{2}{e^{ - {{\left( {\sqrt Q  - \frac{{\left| {\vec c} \right|}}{2}} \right)}^2}}}  - \frac{1}{2}{\left| {\vec c} \right|^2}{e^{ - {{\left( {\sqrt Q  + \frac{{\left| {\vec c} \right|}}{2}} \right)}^2}}} - \frac{1}{2}{\left| {\vec c} \right|^2}{e^{ - {{\left( {\sqrt Q  - \frac{{\left| {\vec c} \right|}}{2}} \right)}^2}}},
\end{aligned}
\]
\[
\begin{aligned}
\overline{I_G}\left( {Q,\vec c} \right) & = \frac{1}{{4\pi }}\int\limits_{{{\left| {\vec w} \right|}^2} > Q} {{{\left| {\vec w} \right|}^3}{e^{ - {{\left| {\vec w - \frac{{\vec c}}{2}} \right|}^2}}}d\vec w}  \\
   & = \frac{{3\sqrt \pi  }}{{16\left| {\vec c} \right|}}\left( {{\rm erf}\left( {\sqrt Q  + \frac{{\left| {\vec c} \right|}}{2}} \right) - {\rm erf}\left( {\sqrt Q  - \frac{{\left| {\vec c} \right|}}{2}} \right)} \right) \\
   & + \frac{{3\sqrt \pi  }}{{16}}\left| {\vec c} \right|\left( {{\rm erf}\left( {\sqrt Q  + \frac{{\left| {\vec c} \right|}}{2}} \right) - erf\left( {\sqrt Q  - \frac{{\left| {\vec c} \right|}}{2}} \right)} \right) \\
   & + \frac{{\sqrt \pi  }}{{64}}{\left| {\vec c} \right|^3}\left( {{\rm erf}\left( {\sqrt Q  + \frac{{\left| {\vec c} \right|}}{2}} \right) - {\rm erf}\left( {\sqrt Q  - \frac{{\left| {\vec c} \right|}}{2}} \right)} \right) \\
   & - \frac{3}{{8\left| {\vec c} \right|}}\left( {\sqrt Q  + \frac{{\left| {\vec c} \right|}}{2}} \right){e^{ - {{\left( {\sqrt Q  + \frac{{\left| {\vec c} \right|}}{2}} \right)}^2}}} + \frac{3}{{8\left| {\vec c} \right|}}\left( {\sqrt Q  - \frac{{\left| {\vec c} \right|}}{2}} \right){e^{ - {{\left( {\sqrt Q  - \frac{{\left| {\vec c} \right|}}{2}} \right)}^2}}} \\
   & - \frac{1}{{4\left| {\vec c} \right|}}{\left( {\sqrt Q  + \frac{{\left| {\vec c} \right|}}{2}} \right)^3}{e^{ - {{\left( {\sqrt Q  + \frac{{\left| {\vec c} \right|}}{2}} \right)}^2}}} + \frac{1}{{4\left| {\vec c} \right|}}{\left( {\sqrt Q  - \frac{{\left| {\vec c} \right|}}{2}} \right)^3}{e^{ - {{\left( {\sqrt Q  - \frac{{\left| {\vec c} \right|}}{2}} \right)}^2}}} \\
   & - \frac{3}{8}\left| {\vec c} \right|\left( {\sqrt Q  + \frac{{\left| {\vec c} \right|}}{2}} \right){e^{ - {{\left( {\sqrt Q  + \frac{{\left| {\vec c} \right|}}{2}} \right)}^2}}} + \frac{3}{8}\left| {\vec c} \right|\left( {\sqrt Q  - \frac{{\left| {\vec c} \right|}}{2}} \right){e^{ - {{\left( {\sqrt Q  - \frac{{\left| {\vec c} \right|}}{2}} \right)}^2}}} \\
   & + \frac{1}{2}{\left( {\sqrt Q  + \frac{{\left| {\vec c} \right|}}{2}} \right)^2}{e^{ - {{\left( {\sqrt Q  + \frac{{\left| {\vec c} \right|}}{2}} \right)}^2}}} + \frac{1}{2}{\left( {\sqrt Q  - \frac{{\left| {\vec c} \right|}}{2}} \right)^2}{e^{ - {{\left( {\sqrt Q  - \frac{{\left| {\vec c} \right|}}{2}} \right)}^2}}} \\
   & + \frac{1}{2}{e^{ - {{\left( {\sqrt Q  + \frac{{\left| {\vec c} \right|}}{2}} \right)}^2}}} + \frac{1}{2}{e^{ - {{\left( {\sqrt Q  - \frac{{\left| {\vec c} \right|}}{2}} \right)}^2}}} + \frac{1}{2}{\left| {\vec c} \right|^2}{e^{ - {{\left( {\sqrt Q  + \frac{{\left| {\vec c} \right|}}{2}} \right)}^2}}} + \frac{1}{2}{\left| {\vec c} \right|^2}{e^{ - {{\left( {\sqrt Q  - \frac{{\left| {\vec c} \right|}}{2}} \right)}^2}}},
\end{aligned}
\]
\[
\begin{aligned}
  {I_H}\left( {Q,\vec c} \right) & = \frac{1}{{2\pi }}\int\limits_{0 < {{\left| {\vec w} \right|}^2} < Q} {\left| \vec w \right| \left( {\vec w,\vec c} \right){e^{ - {{\left| {\vec w - \frac{{\vec c}}{2}} \right|}^2}}}d\vec w}  \\
   & = 4{I_G}\left( {Q,\vec c} \right) - 4{I_F}\left( {Q,\vec c} \right) + \frac{{{Q^{3/2}}}}{{\left| {\vec c} \right|}}\left( {{e^{ - {{\left( {\sqrt Q  - \frac{{\left| {\vec c} \right|}}{2}} \right)}^2}}} - {e^{ - {{\left( {\sqrt Q  + \frac{{\left| {\vec c} \right|}}{2}} \right)}^2}}}} \right),
\end{aligned}
\]
\[
\begin{aligned}
  \overline{I_H}\left( {Q,\vec c} \right) & = \frac{1}{{2\pi }}\int\limits_{{{\left| {\vec w} \right|}^2} > Q} {\left| \vec w \right| \left( {\vec w,\vec c} \right){e^{ - {{\left| {\vec w - \frac{{\vec c}}{2}} \right|}^2}}}d\vec w}  \\
   & = 4\overline{I_G}\left( {Q,\vec c} \right) - 4\overline{I_F}\left( {Q,\vec c} \right) - \frac{{{Q^{3/2}}}}{{\left| {\vec c} \right|}}\left( {{e^{ - {{\left( {\sqrt Q  - \frac{{\left| {\vec c} \right|}}{2}} \right)}^2}}} - {e^{ - {{\left( {\sqrt Q  + \frac{{\left| {\vec c} \right|}}{2}} \right)}^2}}}} \right),
\end{aligned}
\]
\[
\begin{aligned}
  {I_{GH}}\left( {Q,\vec c} \right) & = \frac{1}{{4\pi }}\int\limits_{0 < {{\left| {\vec w} \right|}^2} < Q} {\left| {\vec w} \right|{{\left( \vec w, {\vec w - \frac{{\vec c}}{2}} \right)}}{e^{ - {{\left| {\vec w - \frac{{\vec c}}{2}} \right|}^2}}}d\vec w}  \\
   & = \frac{1}{2} e^{- \frac{| \vec c |^2}{4}} + \frac{\sqrt{\pi}}{2 | \vec c |} \left(1 + \frac{| \vec c |^2}{4} \right) {\rm erf} \left( \frac{| \vec c |}{2} \right) \\
   & - \frac{\sqrt{\pi}}{4 | \vec c |} \left(1 + \frac{| \vec c |^2}{2} \right) \left( {\rm erf} \left( \sqrt{Q} + \frac{| \vec c |}{2} \right) - {\rm erf} \left( \sqrt{Q} - \frac{| \vec c |}{2} \right) \right) \\
   & - \frac{1}{2 | \vec c |} \left( \sqrt{Q} + \frac{|\vec c|}{2} \right) e^{- \left( \sqrt{Q} - \frac{| \vec c |}{2} \right)^2} + \frac{1}{2 | \vec c |} \left( \sqrt{Q} - \frac{|\vec c|}{2} \right) e^{- \left( \sqrt{Q} + \frac{| \vec c |}{2} \right)^2} \\
   & - \frac{{{Q^{3/2}}}}{{4\left| {\vec c} \right|}}\left( {{e^{ - {{\left( {\sqrt Q  - \frac{{\left| {\vec c} \right|}}{2}} \right)}^2}}} - {e^{ - {{\left( {\sqrt Q  + \frac{{\left| {\vec c} \right|}}{2}} \right)}^2}}}} \right),
\end{aligned}
\]
\[
\begin{aligned}
  \overline{I_{GH}}\left( {Q,\vec c} \right) & = \frac{1}{{4\pi }}\int\limits_{{{\left| {\vec w} \right|}^2} > Q} {\left| {\vec w} \right|{{\left( \vec w, {\vec w - \frac{{\vec c}}{2}} \right)}}{e^{ - {{\left| {\vec w - \frac{{\vec c}}{2}} \right|}^2}}}d\vec w}  \\
   & = \frac{\sqrt{\pi}}{4 | \vec c |} \left(1 + \frac{| \vec c |^2}{2} \right) \left( {\rm erf} \left( \sqrt{Q} + \frac{| \vec c |}{2} \right) - {\rm erf} \left( \sqrt{Q} - \frac{| \vec c |}{2} \right) \right) \\
   & + \frac{1}{2 | \vec c |} \left( \sqrt{Q} + \frac{|\vec c|}{2} \right) e^{- \left( \sqrt{Q} - \frac{| \vec c |}{2} \right)^2} - \frac{1}{2 | \vec c |} \left( \sqrt{Q} - \frac{|\vec c|}{2} \right) e^{- \left( \sqrt{Q} + \frac{| \vec c |}{2} \right)^2} \\
   & + \frac{{{Q^{3/2}}}}{{4\left| {\vec c} \right|}}\left( {{e^{ - {{\left( {\sqrt Q  - \frac{{\left| {\vec c} \right|}}{2}} \right)}^2}}} - {e^{ - {{\left( {\sqrt Q  + \frac{{\left| {\vec c} \right|}}{2}} \right)}^2}}}} \right),
\end{aligned}
\]
\[
\begin{aligned}
  {I_{GH2}}\left( {Q,\vec c} \right) & = \frac{1}{{4\pi }}\int\limits_{0 < {{\left| {\vec w} \right|}^2} < Q} {\left| {\vec w} \right|{{\left| {\vec w - \frac{{\vec c}}{2}} \right|}^2}{e^{ - {{\left| {\vec w - \frac{{\vec c}}{2}} \right|}^2}}}d\vec w}  \\
   & = {I_G}\left( {Q,\vec c} \right) - \frac{1}{2}{I_H}\left( {Q,\vec c} \right) + \frac{{{{\left| {\vec c} \right|}^2}}}{8}{I_F}\left( {Q,\vec c} \right) \\
   & = \frac38 e^{- \frac{| \vec c |^2}{4}} + \frac{\sqrt{\pi}}{16 | \vec c |} \left(10 + 3 |\vec c|^2 \right) {\rm erf} \left( \frac{|\vec c|}{2} \right) \\
   & - \frac{\sqrt{\pi}}{32| \vec c |} \left(10 + 3 | \vec c | \right) \left( {\rm erf} \left( \sqrt{Q} + \frac{| \vec c |}{2} \right) - {\rm erf} \left( \sqrt{Q} - \frac{| \vec c |}{2} \right) \right) \\
   & + \frac{\sqrt{Q}}{8 |\vec c|} \left( 2Q + 5 \right) \left( e^{-\left( \sqrt{Q} + \frac{| \vec c |}{2} \right)^2} - e^{-\left( \sqrt{Q} - \frac{| \vec c |}{2} \right)^2} \right) \\
   & + \frac{1}{16} \left( 2Q - 3 \right) \left( e^{-\left( \sqrt{Q} + \frac{| \vec c |}{2} \right)^2} + e^{-\left( \sqrt{Q} - \frac{| \vec c |}{2} \right)^2} \right) \\
  \mathop  \to \limits_{\left| {\vec c} \right| \to 0}  \\
  & 1 - \left( {1 + Q + \frac{{{Q^2}}}{2}} \right){e^{ - Q}},
\end{aligned}
\]
\[
\begin{aligned}
  \overline{I_{GH2}}\left( {Q,\vec c} \right) & = \frac{1}{{4\pi }}\int\limits_{{{\left| {\vec w} \right|}^2} > Q} {\left| {\vec w} \right|{{\left| {\vec w - \frac{{\vec c}}{2}} \right|}^2}{e^{ - {{\left| {\vec w - \frac{{\vec c}}{2}} \right|}^2}}}d\vec w}  \\
   & = \overline{I_G}\left( {Q,\vec c} \right) - \frac{1}{2}\overline{I_H}\left( {Q,\vec c} \right) + \frac{{{{\left| {\vec c} \right|}^2}}}{8}\overline{I_F}\left( {Q,\vec c} \right) \\
   & = \frac{\sqrt{\pi}}{32| \vec c |} \left(10 + 3 | \vec c |^2 \right) \left( {\rm erf} \left( \sqrt{Q} + \frac{| \vec c |}{2} \right) - {\rm erf} \left( \sqrt{Q} - \frac{| \vec c |}{2} \right) \right) \\
   & - \frac{\sqrt{Q}}{8 |\vec c|} \left( 2Q + 5 \right) \left( e^{-\left( \sqrt{Q} + \frac{| \vec c |}{2} \right)^2} - e^{-\left( \sqrt{Q} - \frac{| \vec c |}{2} \right)^2} \right) \\
   & - \frac{1}{16} \left( 2Q - 3 \right) \left( e^{-\left( \sqrt{Q} + \frac{| \vec c |}{2} \right)^2} + e^{-\left( \sqrt{Q} - \frac{| \vec c |}{2} \right)^2} \right).
\end{aligned}
\]
\section{Beyond Euler's hydrodynamics: Grad's 13-moment (14-moment) approach.}

\subsection{Derivation of Grad's 14-moment equations}
The hydrodynamic description of fluids is based on the assumption of two well-separated time scales -- one ("fast"), collision time scale and another -- ("slow") hydrodynamic time scale \cite{Resibua, Garzo}. The first stage defines the so-called \textit{kinetic} regime, sensitive to initial conditions of fluid. The second stage defines the \textit{hydrodynamic} regime where details of the initial conditions are  completely forgotten  and the state is determined by hydrodynamic fields. In the hydrodynamic stage the  dependence of the distribution function on time and space occurs only through hydrodynamic fields -- density $n(\vec{r},t)$, flux velocity $\vec{u}(\vec{r},t)$ and local temperature, $T(\vec{r},t)$, which  correspond to  zeros, first and second-order  moment of the velocity  distribution function (VDF) $f(\vec{V},\vec{r},t)$. The VDF in the hydrodynamic regime depends on time and spatial coordinates only through these (and sometimes other) hydrodynamic fields. In equilibrium fluids  of particles that collide elastically, the VDF is Maxwellian:
\begin{equation}
\label{eq:Max2}
  f \left( \vec V, \vec{r},t \right) = f_M \left( \vec V, \vec{r},t \right)=\frac{n }{\left( 2 \pi \theta \right)^{3/2}} e^{\frac{-\left( \vec V - \vec u \right)^2}{2 \theta}}.
\end{equation}
It contains five first  moments -- zero, first and second-order moments, which are:  $n = n\left(\vec{r},t \right) = \int \!f d \vec{V}$  -- the number density of particles,
$\vec{u}=\vec{u}\left( \vec{r},t \right) = n^{-1} \int \vec{V} f d \vec{V}$ -- the respective flux velocity and  $\theta=T\left( \vec{r},t \right)/m$ -- the reduced  temperature of such particles, $\theta = \tfrac13 n^{-1} \int  (\vec{V} -\vec{u})^2 f d \vec{V}$, with $m$ being the particles' mass. Using only these fields with the  Maxwellian VDF \eqref{eq:Max2} corresponds to the Euler's level of hydrodynamics \cite{Resibua,Gradbook,Garzo}.

If fluid is not in equilibrium, due to space gradients of density, temperature, or some non-zero fluxes exist, the VDF deviates from the Maxwellian, although the deviations are usually small. To account for the deviations from the Maxwellian distribution two main approaches have been developed -- the Chapman-Enskog and Grad's approach \cite{Resibua,Gradbook,Garzo,BrilliantovPoeschelOUP}. The former is based on the small gradient expansion of the hydrodynamic fields, while the latter is more general. Here we will consider Grad's approach \cite{Gradbook}. The main idea there is to approximate a solution of the Boltzmann equation (BE) by a VDF, whose first few moments are equal to those of the true solution of the BE. This is achieved by the expansion of deviations of the VDF from the Maxwellian in a series of orthogonal polynomials -- Hermite polynomials, $H_{i_1,i_2,\ldots,i_n}^{(n)}(\vec{\xi})$,  of the reduced peculiar velocity, $\vec{\xi} = (\vec{V}-\vec{u})/\sqrt{\theta}=\vec{v}/\sqrt{\theta}$  \cite{Resibua,Gradbook,Garzo}:
\begin{widetext}
\begin{equation}\label{Hermexp}
f \left( \vec V, \vec{r},t \right) = f_M \left( \vec V, \vec{r},t \right)\left[\sum_{n=0}^{\infty}\sum_{i_1,i_2,\ldots,i_n} \frac{1}{n!} a_{i_1,i_2,\ldots,i_n}^{(n)}H_{i_1,i_2,\ldots,i_n}^{(n)}\left(\frac{\vec{V}-\vec{u}}{\sqrt{\theta}}\right) \right].
\end{equation}
Since Hermite polynomials are orthogonal, the expansion coefficients $a_{i_1,i_2,\ldots,i_n}^{(n)}$ may be obtained multiplying Eq. \eqref{Hermexp} with $H_{i_1,i_2,\ldots,i_n}^{(n)}(\vec{\xi})$ and integrating over the velocity $\vec{V}$, that is, $a_{i_1,i_2,\ldots,i_n}^{(n)} =\langle  H_{i_1,i_2,\ldots,i_n}^{(n)}(\vec{\xi}) \rangle$, where the averaging is performed with the VDF. 
It may be shown, that the first few Hermite polynomials are associated with the basic hydrodynamic fields -- density $n$, flux velocity $\vec{u}$, temperature $T$, pressure tensor $P_{\al \be}$ and heat flux $\vec{q}$ (13 variables in total); their averages --  which are the respective expansion coefficients, $a_{i_1,i_2,\ldots,i_n}^{(n)}$, correspond to these fields  \cite{Resibua,Gradbook,Garzo}. As a result, the 13-moment Grad's approximation for the VDF is obtained \cite{Garzo,GarzoGrad,Gradbook}:
\bel{Grad13}
f(\vec{V},\vec{r},t)=f_M(\vec{V},\vec{r},t) \left[1+ \frac{m v_{\al}v_{\be}}{2nT^2} \left(P_{\al \be} -p\delta_{\al \be}\right) + \frac25 \frac{m}{nT^2} \vec{S}_q(\vec{v}) \cdot \vec{q} \right],
\ee
where $\vec{v}=\vec{V}- \vec{u}$ and the summation over repeated (Greek) indexes is implied. The stress tensor, $P_{\al\be}$, pressure, $p$, heat flux $\vec{q}$ and vector $\vec{S}_q(\vec{v})$ are defined as
\bel{Pab}
P_{\al\be}=\int  mv_{\al}v_{\be} f(\vec{V}, \vec{r},t)d\vec{V}, \qquad \qquad p=nT, \qquad \qquad \vec{q}= \int \vec{v} \, \frac{m v^2}{2} f(\vec{V}, \vec{r},t) d\vec{V}, \qquad \qquad \vec{S}_q(\vec{v})=\left( \frac{mv^2}{2T}-\frac52 \right) \vec{v}.
\ee
The Grad's 13-moment approach contains moments of the VDF up to the third order, which may be non-zero only in the presence of fluxes or gradients. Hence, it describes the deviation of the VDF from the Maxwellian due to space gradients and fluxes. Additional deviations, due to collision inelasticity, are described by the fourth moment of VDF, which leads to the 14-moment approximation for the VDF \cite{Garzo,GarzoGrad}:
\bel{Grad14copy}
f(\vec{V},\vec{r},t)=f_M(\vec{V},\vec{r},t) \left[1+ \frac{m  v_{\al}v_{\be}}{2nT^2} \left(P_{\al \be} -p\delta_{\al \be}\right) + \frac25 \frac{m}{nT^2}\vec{S}_q(\vec{v}) \cdot \vec{q} + \frac{a_2}{2} R_2(v^2) \right],
\ee
where  $a_2$ is the full contracted moment of the VDF of the fourth order \cite{Garzo,GarzoGrad}:
\bel{a_2copy}
a_2=\frac{8}{15}\left[\frac{m^2}{4nT^2} \int  V^4 f(\vec{v}, \vec{r},t) d\vec{v} -\frac{15}{4} \right],
\ee
while $R_2(v)$ is the second-order Sonine polynomial,

\bel{SEcopy}
R_2(v^2)= \frac12\left(\frac{v}{v_T}\right)^4 -\frac52 \left(\frac{v}{v_T}\right)^2 +\frac{15}{8},
\ee
with the thermal velocity $v_T^2= 2T/m$. Thus, the last term in Eq. \eqref{Grad14copy} describes deviations from the Maxwellian due to inelasticity, in the lack of fluxes and space gradients.

In our case of aggregating particles (which corresponds to  sticking collisions, with zero restitution coefficient, $\varepsilon=0$), we also expect such deviations as for granular gases. This motivates us to use 14-moment expression for the VDF. For the multi-component system which emerges due to merging collisions, the corresponding VDF for clusters of size $k$ reads,

\begin{eqnarray}
\label{Grad14kcopy}
f_k(\vec{V}_k,\vec{r},t) &=& f_{k,M}(\vec{V}_k,\vec{r},t) + f_{k,NM} (\vec{V}_k,\vec{r},t) =
\frac{n_k}{\left(2 \pi \theta_k\right)^{3/2}}e^{-v_k^2/2\theta_k} \\ &+&\frac{n_k}{\left(2 \pi \theta_k\right)^{3/2}}e^{-v_k^2/2\theta_k}\left[\frac{v_{k,\al} v_{k, \be} }{2m_k n_k\theta_k^2} \left(P_{k,\al \be} -p_k\delta_{\al \be}\right)+ \frac{2 }{5 m_k n_k \theta_k^2}\vec{S}_q(\vec{v}_k) \cdot \vec{q}_k + \frac{a_2^{(k)}}{2} R_2(v_k^2) \right] \nonumber,
\end{eqnarray}
which is the sum of the Maxwellian part, $f_{k,M}(\vec{V}_k,\vec{r},t) = n_k/(2\pi \theta_k)^{3/2}\exp\left(-v_k^2/2 \theta_k\right)$ and non-Maxwellian one, $f_{k,NM} (\vec{V}_k,\vec{r},t) $, with the similar definitions of the respective  stress tensor $P_{k,\al \be}$, pressure, $p_k$, heat flux $\vec{q}_k$, reduced temperature, $\theta_k =T_k/m_k$ and coefficient $a_2^{(k)}$.

Once the form of the VDF is determined, one needs to find the fields $P_{k,\al \be}$, $p_k$, $\vec{q}_k$, $T_k$ and $a_2^{(k)}$. These may be found from the moment equations obtained from the BE. Below we consider for simplicity the force-free case when all collisions are merging, $I^{\rm res}=0$, and sources are lacking.

Multiplying Boltzmann equation  (1) with $1$, $m_k \vec{V}_k$ and $m_k v_{k, \al}v_{k,\be}$ and integrating over $\vec{V}_k$ we obtain:
\beel{nk1}
&&\frac{\del n_k}{\del t} + \vec{\nabla} \cdot \left(n_k \vec{u}_k \right) = S_1^{(k)}+ \Delta S_1^{(k)} \\
&&
\label{uk1}
\frac{\partial}{\partial t} \left( m_k n_k  u_{k,\al} \right) +  m_kn_k u_{k,\be} \nabla_{\be} u_{k,\al} + u_{k,\al} \nabla_{\be} \cdot \left( m_kn_k u_{k,\be} \right) + \nabla_{\be} P_{k,\al \be} = m_k{S}_{2,\al}^{(k)}+ m_k\Delta {S}_{2,\al}^{(k)} \\
&&
\label{Pk1}
\frac{\partial}{\partial t} P_{k,\al\be} + P_{k,\al\be} \vec{\nabla}\cdot \vec{u}_k + \vec{u}_k \cdot \vec{\nabla}P_{k,\al\be}+ P_{k,\gamma\be}\nabla_{\gamma}\,u_{k,\al}+
P_{k,\gamma\al}\nabla_{\gamma}\,u_{k,\be }+\frac{2}{5}\nabla_{\gamma}\left( q_{\al}\d_{\be \gamma} + q_{\be}\d_{\al \gamma} + q_{\gamma}\d_{\al \be}\right) \\
&&\qquad \qquad\qquad\qquad\qquad \qquad\qquad\qquad \qquad\qquad\qquad \qquad~~~~~~=  m_kS_{3,\al\be}^{(k)}+ m_k\Delta S_{3,\al\be}^{(k)}. \nn
\eeq
Here we define
\beel{SSS}
S_1^{(k)} \eq \int I^{\rm agg} \left( \{f_{k,M} \}\right) d\vec{V}_k \qquad \qquad \qquad \qquad
\D S_1^{(k)}= \int I^{\rm agg} \left( \{f_{k,M} \}, \{f_{k,NM} \}\right) d\vec{V}_k \qquad \qquad\\
\label{SS2}
\vec{S}_2^{(k)} \eq \int I^{\rm agg} \left( \{f_{k,M} \}\right) \vec{V}_k d\vec{V}_k \qquad \qquad \qquad  \quad
\D \vec{S}_2^{(k)}= \int I^{\rm agg} \left( \{f_{k,M} \}, \{f_{k,NM} \}\right) \vec{V}_k d\vec{V}_k \\
\label{S3}
S_{3,\al\be}^{(k)} \eq \int I^{\rm agg} \left( \{f_{k,M} \}\right) v_{k,\al}v_{k,\be}  d\vec{V}_k \qquad \qquad
\D S_{3,\al\be}^{(k)}= \int I^{\rm agg} \left( \{f_{k,M} \}, \{f_{k,NM} \}\right) v_{k,\al}v_{k,\be} d\vec{V}_k.
\eeq
The notation $(\{f_{k,M} \})$ in the above equations implies that only the Maxwellian part of the VDF is utilized, while $(\{f_{k,M} \},\{f_{k,NM} \})$ -- that both parts of the VDF are used.  Note that the quantities $S_1^{(k)}$,  $\vec{S}_2^{(k)}$ and $S_{3}^{(k)} =S_{3,\al\al}^{(k)}$ have been already defined in  Eqs. (6),  (8) and (9) of the main text. In Eq. \eqref{Pk1} we use the Grad's closure for the third moment of the VDF, expressed in terms of the component of the heat flux $\vec{q}_k$ \cite{Grad,Gradbook}:
\bel{GradClo}
\int m \, v_{k,\al}v_{k,\be} v_{k,\gamma} \, f_k \, d \vec{V}_k = \frac{2}{5}\left( q_{\al}\d_{\be \gamma} + q_{\be}\d_{\al \gamma} + q_{\gamma}\d_{\al \be}\right).
\ee
Multiplying the BE by $\tfrac12 m_k v_k^2 v_{k,\al}$ and integrating over $\vec{V}_k$ we obtain the equation  for the components of $\vec{q}_k$ :
\beel{qeq}
\frac{\del q_{k,\al}}{\del t}&+&\frac32 p_k\frac{\del u_{k,\al}}{\del t}+P_{k,\al \be} \frac{\del u_{k,\be}}{\del t} +\frac52 a_2^{(k)} \d_{\al \be} \nabla_{\be} \left(p_k \theta_k\right) +\frac{1}{n_km_k} \nabla_{\be} \left(\frac54 P_{k,\al \be} -p_k \d_{\al \be} \right) + q_{k,\al} \vec{\nabla} \cdot \vec{u}_k  \\
&+&  \vec{u}_k\cdot \vec{\nabla} q_{k,\al}
 +\frac25 \nabla_{\be} u_{k,\gamma} \left( q_{k,\al}\d_{\be \gamma} +q_{k,\be}\d_{\al \gamma}+ q_{k,\gamma}\d_{\al \be} \right) + \vec{u}_k \cdot \vec{\nabla} u_{k,\be} P_{k, \al \be} +\vec{q}_k \cdot \vec{\nabla } u_{k,\al} \nn \\
 &+&\frac32 p_k \vec{u}_k \cdot \vec{\nabla} u_{k,\al} = m_kS_{4,\al}^{(k)} + m_k \D S_{4,\al}^{(k)},  \nn
\eeq
where we apply similar definitions,
\bel{S4}
S_{4,\al}^{(k)} = \int I^{\rm agg} \left( \{f_{k,M} \}\right) \tfrac12 v_k^2 v_{k,\al}  d\vec{V}_k \qquad \qquad
\D S_{4,\al}^{(k)} = \int I^{\rm agg} \left( \{f_{k,M} \}, \{f_{k,NM} \}\right) \tfrac12  v_k^2 v_{k,\al}  d\vec{V}_k,
\ee
and use Eq. \eqref{GradClo} with  the relation corresponding the Grad's closer \cite{Garzo,GarzoGrad}:
\bel{GradClo2}
\int \tfrac12 m_k v_k^2 v_{k,\al}v_{k,\be} = \frac54 p_k \theta_k a_2^{(k)} \d_{\al \be}+ \frac{p_k}{n_k m_k }\left(\frac54 P_{k, \al \be} -p_k \d_{\al \be} \right) .
\ee
Finally, multiplying the BE by $v_k^4$ and integrating over $\vec{V}_k$ we obtain the equation for  the coefficient $a_2^{(k)}$, characterizing the deviation of the fourth-order moment from that of the Maxwellian VDF:
\bel{a2k}
15 \left( 1 +   a_2^{(k)} \right)m_k^{-1} \left[ \frac{\del (p_k \theta_k)}{\del t} + \frac73 p_k\theta_k \vec{\nabla}\cdot \vec{u}_k + \vec{u}_k \cdot \vec{\nabla} (p_k\theta_k)\right] = S_5^{(k)}+ \D S_5^{(k)},
\ee
with
\bel{SSS5}
S_5^{(k)} = \int I^{\rm agg} \left( \{f_{k,M} \}\right) v_k^4 d\vec{V}_k \qquad \qquad \qquad \qquad
\D S_5^{(k)}= \int I^{\rm agg} \left( \{f_{k,M} \}, \{f_{k,NM} \}\right) v_k^4 d\vec{V}_k.
\ee
Eqs. \eqref{nk1}, \eqref{uk1}, \eqref{Pk1}, \eqref{qeq}, \eqref{a2k} along with the relations $p_k=m_k n_k\theta_k$, $p_k = \tfrac13 P_{k,\al \al}$, and Eqs. \eqref{SSS},  \eqref{SS2}, \eqref{S3}, \eqref{S4}, \eqref{SSS5} for $S_1^{(k)}-S_5^{(k)}$ and $\D S_1^{(k)}-\D S_5^{(k)}$ form a closed set of equations for 14-moment Grad hydrodynamics; in Eqs. \eqref{SSS}, \eqref{SS2},\eqref{S3}, \eqref{S4}, \eqref{SSS5} one needs to use VDF \eqref{Grad14kcopy}.   Note that with the VDF defined by  \eqref{Grad14kcopy}, the quantities $S_1^{(k)} +\D S_1^{(k)}$ may be written in the following form,

\beel{SiDSi}
S_1^{(k)}+ \D S_1^{(k)}\eq  \frac12\sum_{i+j=k}C_{ij}n_in_j-\sum_{j=1}^{\infty}C_{kj}n_in_j \\
&+&  \frac12\sum_{i+j=k}\left[ \left( C_{ij}^{(p,\al\be)} \Pi_{i,\al\be} + C_{ji}^{(p,\al\be)} \Pi_{j,\al\be}\right) +  \left( C_{ij}^{(q,\gamma )}q_{i, \gamma}+
C_{ji}^{(q,\gamma )}q_{j, \gamma}\right)+ C_{ij}^{(a_2)} (a_2^{(i)} + a_2^{(j)}) \right] n_in_j \nn \\
&-&\sum_{j=1}^{\infty} \left[\left( C_{kj}^{(p,\al\be)} \Pi_{k,\al\be} + C_{jk}^{(p,\al\be)} \Pi_{j,\al\be}\right) +  \left( C_{kj}^{(q,\gamma )}q_{k, \gamma}+
C_{jk}^{(q,\gamma )}q_{j, \gamma}\right)+C_{kj}^{(a_2)} (a_2^{(k)} + a_2^{(j)})\right]n_kn_j . \nn
\eeq
Here the coefficients $C_{ij}$ are defined in Eq. (7) of the main text and correspond to the 5-moment Grad's approximation. In Eq. \eqref{SiDSi}  we abbreviate, $\Pi_{i,\al\be}=P_{i,\al\be}-p_{i}\d_{\al\be}$, and keep only terms, linear with respect to $\Pi_{i,\al\be}$, and coefficients $a_2^{(k)}$; the summation over repeated Greek indexes is implied. All other quantities, $S_2^{(k)}+\D S_2^{(k)}$, $S_3^{(k)}+ \D S_3^{(k)}$,  $S_4^{(k)}+\D S_4^{(k)}$ and $S_5^{(k)}+ \D S_5^{(k)}$ may be written in the similar form. That is,  they contain the basic terms associated with the Maxwellian VDF, which correspond to the 5-moment Grad's approximation (terms with the coefficients $\vec{P}_{ij}$, $\vec{R}_{ij}$, $B_{ij}$ and $D_{ij}$ in Eqs. (8) and (9) of the main text, where $B_{ij}$ and $D_{ij}$ refer to the diagonal part of $S_{3,\al \al}^{(k)}$), and the terms associated with the next 9 moments of the 14-moment Grad's approximation. The latter contains terms associated with the traceless part of the stress tensor, $\Pi_{i,\al\be}$, heat flux, $\vec{q}^{\,(i)}$ and the fourth moment coefficient, $a_2^{(i)}$. All these coefficients may be found explicitly, as they correspond to  Gaussian integrals. As an example we present below the coefficients $C_{ij}^{(p,\al,\be)}$:

\beel{Cijpab}
C_{ij}^{(p,\al\be)}=\pi \sigma_{ij}^2 \sqrt{\theta_{ij}} \left[\frac{\omega_{0,ij} \theta_{j} +\omega_{1,ij} \theta_{ij}}{n_im_i\theta_i \theta_{ij}}\, \d_{\al \be} +\omega_{2,ij} \,c_{ij,\al} c_{ij,\be} \right],
\eeq
where $\vec{c}_{ij}$ and $\theta_{ij}$ are defined as
\beel{cG}
\vec{c}_{ij} \eq \sqrt{\frac{2}{\theta_{ij}}} (\vec{u}_{i}-\vec{u}_{j}) \\
\theta_{ij} \eq \theta_i +\theta_j
\eeq
and we introduce the coefficients:
\beel{omega}
\omega_{0,ij}  \eq \pi e^{-c_{ij}^2/4} + \frac{\pi^{3/2}(c_{ij}^2+2) {\rm erf}\left(\frac{c_{ij}}{2}\right)}{2c_{ij}}\\
\omega_{1,ij} \eq \frac{1}{2 c_{ij}^3}\left[ \pi^{3/2} c_{ij}^2 (c_{ij}^2+6) -4 \pi c_{ij}e^{-c_{ij}^2/4} -2\pi^{3/2}(c_{ij}^2+2) {\rm erf}\left(\frac{c_{ij}}{2}\right)\right]   \\
\omega_{2,ij}\eq \frac{e^{-c_{ij}^2/4}}{2 c_{ij}^5}\!\left[ 2 \pi c_{ij} (c_{ij}^4+c_{ij}^2+6) \!+\!\pi^{3/2} e^{c_{ij}^2/4}\!\left( c_{ij}^2(c_{ij}^2+6)(2c_{ij}^2-3)+(c_{ij}^6+3c_{ij}^4+4c_{ij}^2+12) {\rm erf}\left(\frac{c_{ij}}{2}\right) \right) \!\right],
\eeq
where $c_{ij}=|\vec{c}_{ij}|$.
\subsection{Euler's and Navier-Stokes hydrodynamics }

Let us analyze the impact of the field gradients. In the linear with respect to gradients approximation, the stress tensor and heat flux read \cite{Resibua,BrilliantovPoeschelOUP,Gradbook}:
\beel{PqNS}
P_{k, \al \be} \eq p_k\d_{\al \be}-\eta_k\left(\nabla_{\al}u_{k,\be} + \nabla_{\be}u_{k,\al}-\frac23 \delta_{\al \be} \vec{\nabla} \cdot \vec{u}_k \right)
\\
\label{PqNS1}
\vec{q}_k \eq -\kappa_k \vec{\nabla}T_k -\mu_k \vec{\nabla}n_k,
\eeq
where $\eta_k$ is the viscosity associated with the component of particles of size $k$, $\kappa_k$ is the respective thermal conductivity of this component and $\mu_k$ is the new transport coefficient which appear in granular gases with inelastic collisions, see e.g. \cite{BrilliantovPoeschelOUP}. We expect the appearance of such a coefficient in the aggregating gases as well. The linear expressions for $P_{k,\al\be}$ and $\vec{q}_k$ in terms of the field gradients correspond to the Navier-Stokes hydrodynamics.

The simplest approximation is the five-moments approximation, when only $n_k$, $\vec{u}_k=(u_{k,x},u_{k,y},u_{k,z})$ and $T_k$ are used; it  is associated with the Maxwellian VDF, Eq. \eqref{Grad14kcopy}. In this case the non-diagonal elements of $P_{k,\al \be}$, heat flux and coefficient $a_2^{(k)}$ are neglected, that is $P_{k,\al \be}=m_kn_k\theta_k\d_{k,\al \be}$, $\vec{q}_k=0$ and $a_2^{(k)}=0.$ Hence we obtain for Eqs. \eqref{nk1} and \eqref{uk1}:
\beel{nk2}
&&\frac{\del n_k}{\del t} + \vec{\nabla} \cdot \left(n_k \vec{u}_k \right) = S_1^{(k)}, \\
&&
\label{uk2}
\frac{\partial}{\partial t} \left( n_k  \vec{u}_{k} \right) +  n_k \vec{u}_{k} \cdot \vec{\nabla} \vec{u}_k + \vec{u}_{k} \vec{\nabla} \cdot(n_k \vec{u}_k ) + \vec{\nabla} \left( n_k \theta_k \right) = \vec{S}_2^{(k)},
\eeq
which coincide with Eqs. (6) and (8) of the main text. In the five-moment Grad's approximation  only diagonal components in Eq. \eqref{Pk1} are non-zero. We sum the diagonal components of this equation and multiply the result by $(3m_k)^{-1}$ to obtain:
\bel{Pk2}
\frac{\partial}{\partial t} \left( n_k \theta_k \right)+ \theta_k \vec{\nabla} \cdot (n_k \vec{u}_k) + n_k\vec{u}_k \cdot \vec{\nabla} \theta_k +\frac23 n_k \theta_k \vec{\nabla} \cdot \vec{u}_k = S_3^{(k)},
\ee
which coincides with Eq. (9) of the main text.

Let us consider the derivation of transport coefficients for the Navier-Stokes level of hydrodynamic description. Here we address  the case of  viscosity coefficients $\eta_k$ and assume that temperature and density gradients are lacking, $\vec{\nabla} T_k=0$, $\vec{\nabla} n_k=0$; we also neglect the coefficient $a_2^{(k)}$, i.e. we assume that  $a_2^{(k)}=0$. Consider then the non-diagonal part of Eq. \eqref{Pk1}. Using Eq. \eqref{PqNS} for $P_{k,\al \be}$, we obtain the equations for the viscosity coefficients $\eta_i$ ($\al \neq \be$):
\beel{visc}
\frac{\del}{\del t} \eta_k R_{k,\al \be} = {\cal A}_{k,\al \be} &+& \frac12\sum_{i+j=k} \left[ \eta_i {\cal D}_{ki, \al \be,\al_1\be_1 }R_{i,\al_1\be_1}+ \eta_j {\cal D}_{kj,\al \be,\al_2\be_2 }R_{i,\al_2\be_2} \right] \\
&-&
\sum_{j=1}^{\infty} \left[ \eta_k {\cal B}_{kk, \al \be,\al_3\be_3 }R_{k,\al_3\be_3}+ \eta_j {\cal B}_{kj,\al \be,\al_4\be_4 }R_{j,\al_4\be_4} \right] \nn,
\eeq
where the summation over repeated Greek indexes is implied,
\bel{Rk}
R_{k, \al \be }= \nabla_{\al}u_{k,\be} + \nabla_{\be}u_{k,\al},
\ee
and we define the tensorial coefficients:
\beel{AAk}
{\cal A}_{k,\al \be} \eq \frac12 \sum_{i+j=k} \sigma_{ij}^2\int d\vec{V}_kd\vec{V}_id\vec{V}_jd\vec{e} v_{k,\al}v_{k,\be} |\vec{V}_{ij}\cdot \vec{e}| \theta (-\vec{V}_{ij} \cdot \vec{e}) \d \left( \left(m_i + m_j \right) \vec{V}_k-m_i \vec{V}_i -m_j \vec{V}_j \right) f_{i,M}f_{j,M} \\
&-&\sum_{j=1}^{\infty} \sigma_{kj}^2\int d\vec{V}_kd\vec{V}_jd\vec{e} v_{k,\al}v_{k,\be} |\vec{V}_{kj}\cdot \vec{e}| \theta (-\vec{V}_{kj} \cdot \vec{e}) f_{k,M}f_{j,M}. \nn
\eeq
and
\beel{DDDk}
{\cal D}_{ki,\al\be,\al_1\be_1} \eq \sigma_{ij}^2 \int d\vec{V}_kd\vec{V}_id\vec{V}_jd\vec{e} \, \frac{1}{2n_i m_i\theta_i^2} v_{k,\al}v_{k,\be} v_{i,\al_1}v_{i,\be_1}|\vec{V}_{ij}| \theta (-\vec{V}_{ij} \cdot \vec{e}) \d (M_{ij} \vec{V}_k-m_i \vec{V}_i -m_j \vec{V}_j)  f_{i,M}f_{j,M} \\
{\cal B}_{kj,\al\be,\al_1\be_1} \eq  \sigma_{kj}^2\int d\vec{V}_k d\vec{V}_jd\vec{e} \, \frac{1}{2n_j m_j\theta_j^2} v_{k,\al}v_{k,\be} v_{j,\al_1}v_{j,\be_1}|\vec{V}_{kj}| \theta (-\vec{V}_{kj} \cdot \vec{e})  \vec{V}_{ij}f_{k,M}f_{j,M},
\eeq
where, as previously, $\vec{V}_{ij}=\vec{V}_{i}-\vec{V}_{j}$ and
$f_{k,M} = f_{k,M} \left( \vec V_k, t \right)$ denotes the Maxwellian VDF for clusters of size $k$. In principle, one can find the explicit expressions for all the above coefficients ${\cal A}$,  ${\cal B}$, ${\cal D}$, since they may be reduced to the Gaussian integrals. The resulting expressions are however very cumbersome. For instance, the coefficients ${\cal A}_{k,\al \be}$ read:
\beel{Akab}
{\cal A}_{k,\al \be} \eq \frac12\sum_{i+j=k}  \sqrt{\frac{2}{\pi}} n_in_j \sigma_{ij}^2 \sqrt{\theta_{ij}} \left[\left(\omega_{0,ij}\frac{\theta_i \theta_j}{\theta_{ij}}+\omega_{1,ij}\frac{(m_i\theta_i-m_j \theta_j)^2}{(m_i+m_j)^2\theta_{ij}}\right) \d_{\al \be} +\omega_{2,ij} \frac{(m_i\theta_i-m_j \theta_j)^2}{(m_i+m_j)^2\theta_{ij}} c_{ij,\al}c_{ij,\be} \right]\\
&-& \sum_{j=1}^{\infty} \sqrt{\frac{2}{\pi}} n_kn_j \sigma_{kj}^2 \sqrt{\theta_{kj}} \left[\left(\omega_{0,kj}\frac{\theta_k \theta_j}{\theta_{kj}}+\omega_{1,kj}\frac{\theta_k^2}{\theta_{kj}}\right) \d_{\al \be} +\omega_{2,kj}\frac{\theta_k^2}{\theta_{kj}} c_{kj,\al}c_{kj,\be} \right], \nn
\eeq
where all the notations, along with the coefficients $\omega_{l,ij}$, with $l=0,1,2$ have been introduced above.
The coefficients ${\cal D}_{ki,\al\be,\al_1\be_1}$ and ${\cal B}_{ki,\al\be,\al_1\be_1}$ have even more complicated expressions.
\end{widetext}
As it follows from the above Eqs. \eqref{visc} the viscosity coefficients which may be found  from these equations, generally depend on the flux velocity gradients $\nabla_{\al} u_{k, \be} $, which implies non-Newtonian viscous behavior of the fluid. That is, we conclude that aggregation processes, which take place in fluids, make their behavior non-Newtonian.

Similar analysis may be performed for the heat conductivity coefficients $\kappa_k$ and coefficients $a_2^{(k)}$. The resulting expressions are also very complicated. Note that in Eqs. \eqref{PqNS}, \eqref{PqNS1} we use the simplest form for the non-diagonal part of the stress tensor and the heat flux. One can use a more general form, $\Pi_{k,\al\be} = \sum_{j} \eta_{jk} R_{j,\al \be}$ and $\vec{q}_k= -\sum_j (\kappa_{kj} \vec{\nabla}T_j +\mu_{kj} \vec{\nabla}n_j)$ and perform a similar analysis. The qualitative conclusion will be however the same -- fluids with aggregation demonstrate non-Newtonian viscosity.

Hence we conclude that the Grad's 14-hydrodynamics and its Navier-Stokes part based on Eqs. \eqref{PqNS},  \eqref{PqNS1} are too cumbersome  for practical applications. Fortunately, as shown in the main text, the Euler's level of hydrodynamic  description is relatively simple and  demonstrates an acceptable accuracy. Hence it may be successfully used in applications.

\section{Details of the numerical scheme}
Consider now the numerical analysis of the system \eqref{eq:n2}-\eqref{eq:t2}. For simplicity, we address the case, where the non-uniformity comes from the $z$-axis, for example, due to the presence of the gravitational force.   Then the equations take the form

\begin{widetext}
\[
\begin{aligned}
 & \frac{\partial}{{\partial t}}{n_k} + \frac{\partial}{\partial z} \left( {n_k \vec u_k} \right) = \frac{1}{2}\sum\limits_{i + j = k} {{C_{ij}}{n_i}{n_j}}  - \sum\limits_{j = 1}^\infty  {{C_{kj}}{n_k}{n_j}} ,\quad k = \overline {1,\infty }, \\
 & \frac{\partial}{{\partial t}} \left( {n_k}{u_k} \right) + \frac{\partial}{\partial z} \left( n_k u_k^2 \right) + \frac{\partial}{\partial z} \left( {n_k \theta_k } \right) + n_k g = \frac{1}{2}\sum\limits_{i + j = k} {{P_{ij}}{n_i}{n_j}}  - \sum\limits_{j = 1}^\infty  {{R_{kj}}{n_k}{n_j}} ,\quad k = \overline {1,\infty } , \\
 & \frac{\partial}{{\partial t}} \left({n_k}{\theta_k} \right) + \frac{\partial}{\partial z} \left( n_k u_k \theta_k \right) + 2 n_k \theta_k \frac{\partial}{\partial z} u_k = \frac{1}{2}\sum\limits_{i + j = k} {{B_{ij}}{n_i}{n_j}}  - \sum\limits_{j = 1}^\infty  {{D_{kj}}{n_k}{n_j}} ,\quad k = \overline {1,\infty } ,
\end{aligned}
\]
We can use space discretization in the vertical direction and apply the appropriate boundary conditions. For example, in the case when particles are falling we should either define the number density, speed and temperatures at the highest point or put an appropriate source there. We should also limit ourselves to a finite number of equations $N$. After that, we can utilize any appropriate time and space discretization. In the simplest case, we can exploit the forward Euler scheme. Note that we can use a smaller time step or a different scheme for the Smoluchowski equations subsystem (decompose equations into parts describing different physical processes) if the aggregation process requires a more accurate consideration.
\end{widetext}

Hence, the simplest time and space discretization reads:
\begin{widetext}
\[
\begin{aligned}
  &n_k \left( t = t_{n+1}, z = h_m \right) = n_k \left( t = t_n, z = h_m \right) \\
 & + \left( \left( n_k u_k \right) \left( t = t_n, z = h_{m-1} \right) - \left( n_k u_k \right) \left( t = t_n, z = h_m \right) \right) \Delta t_n / h \\
 & + \left( \frac{1}{2}\sum\limits_{i + j = k} {{C_{ij}}{n_i}{n_j}}  - \sum\limits_{j = 1}^\infty  {{C_{kj} (t_n)}{n_k}{n_j}} \right) \left( t = t_n, z = h_m \right) \Delta t_n ,\quad k = \overline {1,\infty }, \\
 & \left( n_k u_k \right) \left( t = t_{n+1}, z = h_m \right) = \left( n_k u_k \right) \left( t = t_n, z = h_m \right) - n_k \left( t = t_n, z = h_m \right) g \Delta t_n \\
 & + \left( \left( n_k u_k^2 + n_k \theta_k \right) \left( t = t_n, z = h_{m-1} \right) - \left( n_k u_k^2 + n_k \theta_k \right) \left( t = t_n, z = h_m \right) \right) \Delta t_n / h \\
 & + \left(\frac{1}{2}\sum\limits_{i + j = k} {{P_{ij}}{n_i}{n_j}}  - \sum\limits_{j = 1}^\infty  {{R_{kj}}{n_k}{n_j}} \right) \left( t = t_n, z = h_m \right) \Delta t_n ,\quad k = \overline {1,\infty } , \\
 & \left( n_k \theta_k \right) \left( t = t_{n+1}, z = h_m \right) = \left( n_k \theta_k \right) \left( t = t_n, z = h_m \right) \\
 & + \left( \left( n_k u_k \theta_k \right) \left( t = t_n, z = h_{m-1} \right) - \left( n_k u_k \theta_k \right) \left( t = t_n, z = h_m \right) \right) \Delta t_n / h \\
 & + \left(\frac{1}{2}\sum\limits_{i + j = k} {{B_{ij}}{n_i}{n_j}}  - \sum\limits_{j = 1}^\infty  {{D_{kj}}{n_k}{n_j}} \right) \left( t = t_n, z = h_m \right) \Delta t_n ,\quad k = \overline {1,\infty } , \\
 & h_m = -mh, m = \overline{0, m_{\max}}, \quad t_{n+1} = t_n + \Delta t_n.
\end{aligned}
\]
We then used the predictor-corrector approach to have second-order time discretization of the aggregation part as in \cite{gen-smol}, where one can find other details for the solution of temperature-dependent Smoluchowski equations. The time step was selected adaptively to be the minimal between the time step, which guarantees stability of temperature-dependent Smoluchowski equations and $\Delta t (t_n) \leqslant 0.2 \Delta h / \max_{k,m} \left| u_k \left( t_n, h_m \right) \right|$, which guarantees the stability of the vertical transport modelling. Unfortunately, we cannot generally use a low-rank approximation approach, see e.g. \cite{gen-smol}, since, when the flow speed difference is large, the elements of $C_{ij}$ do not change smoothly.

\end{widetext}
\end{widetext}
\section{Direct Simulation Monte Carlo of space-inhomogeneous  systems with fluxes and aggregation}
Here we briefly describe the atomistic simulation method, called ``Direct Simulation Monte Carlo'' (DSMC), namely, its modification for space-inhomogeneous aggregating systems with fluxes. Physically, this method  corresponds to the solution of the respective Boltzmann equation. Hence, it is applicable for systems that are not dense, where the dominating part of inter-particle collisions are pairwise and   successive collisions are not correlated.  DSMC  has been initially proposed by Bird \cite{dsmc} for molecular gases. Later, various generalizations of the method appeared, including its application to dense gases (Enskog equation),  granular gases, e.g. \cite{Hopkins:1992,Montanero:1996,Montanero:1997,Montanero:1999}, and aggregating gases \cite{nature,kalinov2021direct,gen-smol,OsBrilPRE2022}.

The main advantage of the DSMC, as compared to other atomistic simulations, like, e.g. Molecular Dynamics (MD) simulations, is the possibility to deal with a very large number of particles -- by orders of magnitude larger than in other methods.  This is very important for aggregating systems, since (i) the system rapidly becomes multispecies, as the aggregates of different sizes emerge and (ii) the total number of particles quickly decays due to their merging.  Hence, to obtain reliable statistics for each species (aggregates of different  sizes) the initial number of particles (e.g. monomers) should be very large. The DSMC method elaborated for aggregating systems \cite{nature,OsBrilPRE2022} allows for sustaining a required number of particles of different sizes, by applying special tricks like duplication. The problem, however,  becomes much more challenging if one needs to model space-inhomogeneous systems. In this case, all the relevant statistics are required for each space point (for each cell of a  space mesh, in practice). Fortunately, it is possible to overcome the above difficulty, noticing that in aggregating systems, the velocity distribution function for various species (clusters of different sizes) is very close to the Maxwellian; some deviation exists only in a narrow velocity range, around zero velocity (see the next Section VIII).

\subsection{Classical DSMC -- generalization for space-inhomogeneous systems with aggregation}

First, we start from the classical DSMC \cite{dsmc}, generalized  for the case of aggregation \cite{nature} and further generalize it taking into account fluxes  and space inhomogeneity. Let us briefly sketch the main steps of this version of DSMC, which are the following:

\begin{enumerate}
    \item Take $t=t_c$ and advance time $t \to  t+\delta \tau$, by the time interval $\delta \tau$ as in \cite{dsmc}. That is, choose the interval $\delta \tau$ from an exponential distribution, inversely proportional to the maximum possible collision rate (i.e. the number of collisions per unit time in unit volume) between particles in the system.
    \item Select a pair of particles $i$ and $j$ of size $m_i$ and $m_j$ with the probability, proportional to their collision rate upper bound $n_in_j\sigma_{ij} \left( \left| \vec v_i \right| + \left| \vec v_j \right| \right)$.
    \item Accept a collision with the probability equal to the ratio of the correct collision rate $n_in_j\sigma_{ij} \left| \left( \vec v_j - \vec v_i, \vec e \right) \right|$ with random collision direction $\vec e$ and the bound $n_in_j\sigma_{ij} \left( \left| \vec v_i \right| + \left| \vec v_j \right| \right)$.
    \item Perform aggregation, changing the number of clusters as: $N_i := N_i - 1$, $N_j := N_j - 1$, $N_{i+j} := N_{i+j}+1$.
    \item Velocity of the new cluster after aggregation is found from the moment conservation as $\vec v_{i+j} = \frac{m_i \vec v_i + m_j \vec v_i}{m_i+m_j}$.
\end{enumerate}

Here we do not need to keep track of the temperatures or average velocities --  they can be calculated directly after the simulation ends.

\begin{itemize}
\item[6.] If  time $t$, advanced by  several collisional steps, satisfies the condition $t-t_c< \Delta t$, continue with the above collisional steps $1$-$5$, otherwise switch to the transport step, described below.
\end{itemize}

The transport step here is very simple: since we assume a uniform distribution in computation cells, each cluster $i$ is moved to the next cell in the appropriate direction, with the probability proportional to the cluster velocity $v_i$. Namely, in the setup of  vertically falling clusters, each cluster $i$ is moved to the next layer with the probability equal to $v_i \Delta t / h$, where $h$ is the height of each layer. Similar transport steps are applied for the case of radial explosion.

Note that although DSMC is computationally very efficient, it is still extremely difficult to model aggregation in space inhomogeneous   systems. Indeed, to obtain an acceptable accuracy for the VDF of some species, one need to have a large amount of such particles. Furthermore, to model size distribution, one needs to have a large amount of particles for each size. If one deals with a space inhomogeneous system, this amount of particles is needed for each space cell. This makes simulations very challenging, therefore we have elaborated a more effective version of DSMC discussed below.

\subsection{DSMC with  the Maxwellian VDF}

In addition to the classical DSMC, we performed simulations using a modified DSMC with the Maxwellian VDF.  This was justified by our observation that the actual VDF was rather close to the Maxwellian one. Based on this observation we elaborated an efficient version of DSMC, dealing  with space-inhomogeneous  aggregating systems with fluxes. Physically, the main ideas  of the DSMC for such systems remain the same as in \cite{nature},  although one exploits an additional important assumption, that the \textit{local} VDF may be well approximated by the Maxwellian. This assumption, as demonstrated in the next Section, is well justified.

Once the Maxwell distribution is assumed, we no longer need to keep track of the velocities of individual particles for  the respective ensembles. Still, we  need to keep track of partial temperatures, $T_i$, and flow velocities, $\vec u_i$, for these ensembles,  which define velocity distributions for each particular size.  This drastically differs from the approaches  used for space-homogeneous systems in the lack of fluxes, see e.g. \cite{inverse,monte-random}.

The modified DSMC approach comprises, as before,  ``collisional'' and ``transport'' steps and may be formulated as follows. Let  $N_i$ be the number of particles of size $i$ in a chosen cell, and the current time be  $t_c$, then DSMC makes the following  ``collisional'' steps:

\begin{enumerate}
    \item Take $t=t_c$ and advance time $t \to  t+\delta \tau$, by the time interval $\delta \tau$, chosen in accordance with the standard DSMC technique, using the  basic rule of the  acceptance-rejection method, see e.g.  \cite{monte-random}. That is, choose the interval $\delta \tau$ from an exponential distribution, inversely proportional to the maximum possible collision rate between particles.
    \item Select a pair of particles of size $i$ and $j$ with the probability, proportional to $C_{ij} N_i N_j$.
    \item Accept a collision with the probability equal to the ratio of the current collision rate $C_{ij}$ and maximum collision rate.
    \item Perform aggregation, according to the rule: $N_i := N_i - 1$, $N_j := N_j - 1$, $N_{i+j} := N_{i+j}+1$.
    \item Update the partial flux velocities and temperatures, $\vec u_i$, $\vec u_j$, $\vec u_{i+j}$, $T_i$, $T_j$ and $T_{i+j}$, according to the procedure detailed below.
\end{enumerate}
The updated values of $\vec u_l$ and $T_l$, where $l=i,j, (i+j)$, which alter due to the collision of  particles of size $i$ and $j$,  are to be calculated by averaging over all possible pairs of velocities of such particles. This may be done either directly, sampling particles' velocities $\vec{V}_i$ and $\vec{V}_j$, from the respective Maxwellian  distributions, or analytically, performing the appropriate integration. Here we use the latter approach, which yields explicit expressions for the post-collision values $\vec{u}_l^{\,\prime}$ and $T_l'$, where $l=i,j,k$, with $k=i+j$. More detail  about this method may be found, e.g.,  in Ref. \cite{OsBrilPRE2022}, where it was applied for space homogeneous systems in the lack of fluxes.

In the quasi-1D case, the expressions for
the post-collision values $\vec{u}_l^{\,\prime}$ and $T_l'$ in terms of the pre-collision ones read:
\[
\begin{aligned}
  T_i' & = T_i - \\
  & - \frac{T_i^2/i}{\left( T_i/i+T_j/j \right) \left( 1 + \sqrt{\pi}q e^{q^2}{\rm erf}(q)  \right) \left( N_i-1 \right)}, \\
  T_j' & = T_j - \\
  & - \frac{T_j^2/j}{\left( T_i/i+T_j/j \right) \left( 1 + \sqrt{\pi}q e^{q^2} {\rm erf}(q) \right) \left( N_j-1 \right)}, \\
  T_k' & = \frac{N_k T_k + (i+j)w^2 + \left( iT_i+jT_j \right)/ \left( i+j \right)}{N_k + 1} \\
       & + \frac{w \sqrt{2 \pi}(T_i-T_j) {\rm erf}(q)}{\sqrt{T_i/i+T_j/j} \left(e^{-q^2} + \sqrt{\pi} q {\rm erf}(q) \right) \left( N_k + 1 \right)} \\
       & + \frac{\left( T_i-T_j \right)^2 / (i+j)}{\left( T_i/i+T_j/j \right) \left(1 + \sqrt{\pi} q e^{q^2} {\rm erf}(q) \right) \left( N_k + 1 \right)}, \\
  u_i' & = u_i - \\
  & - \frac{\sqrt{\pi/2} T_i {\rm erf}(q)}{i\sqrt{T_i/i+T_j/j} \left(e^{-q^2} + \sqrt{\pi}q{\rm erf}(q) \right) \left( N_i-1 \right)}, \\
  u_j' & = u_j + \\
  & + \frac{\sqrt{\pi/2} T_j {\rm erf}(q)}{j\sqrt{T_i/i+T_j/j} \left(e^{-q^2} + \sqrt{\pi}q{\rm erf}(q) \right) \left( N_j-1 \right)}, \\
  u_k' & = \frac{N_k u_k + (i u_i+j u_j)/(i+j)}{N_k + 1} \\
       & + \frac{\sqrt{\pi/2} \left(T_i-T_j \right) {\rm erf}(q) / (i+j)}{\sqrt{T_i/i+T_j/j}/(e^{-q^2} + \sqrt{\pi} q {\rm erf}(q))(N_k + 1)},
\end{aligned}
\]
where
\[
\begin{aligned}
  w & = \frac{N_k \left( i u_i+j u_j \right)/(i+j) - N_k u_k}{N_k + 1} \\
    & - \frac{\sqrt{\pi/2} \left( T_i-T_j \right) {\rm erf} (q) / (i+j)}{\sqrt{T_i/i+T_j/j} \left(e^{-q^2} + \sqrt{\pi} q {\rm erf}(q) \right) \left(N_k + 1 \right)}, \\
  q & = \frac{ui-uj}{\sqrt{2 (T_i/i+T_j/j)}}.
\end{aligned}
\]
\begin{itemize}
\item[6.] If  time $t$, advanced by  several collisional steps, satisfies the condition $t-t_c< \Delta t$, continue with the above collisional steps $1$-$5$, otherwise switch to the transport step, described below.
\end{itemize}
Such collision steps are to be performed separately for  each cell (vertical cells or radial layers in the respective setups in our study) which, computationally, may be done in parallel.

After the collision steps are completed for all cells,  time is advanced by $\Delta t$, that is, $t_c \to t_c +\Delta t$, and the transport step of the same duration, $\Delta t$, starts.  At the transport step, particles in all cells are transported as follows. For each size $i$, in each cell,  we select the number of particles, proportional to $N_i u_i \Delta t$ (stochastically rounded to the nearest integer) and move them to the next cell in the appropriate direction (i.e., vertically or radially in the respective setups). Since these particles also transport momentum and kinetic energy, we  update accordingly $\vec{u}_i$ and $T_i$ in the involved cells. After completing the transport step, we return to the collisional steps.

{\color{blue}

}

\section{Closeness of the velocity distribution to the Maxwell distribution}
\begin{figure}[!ht]
\center
\begin{subfigure}[t]{0.45\textwidth}
\centering
\includegraphics[width = \columnwidth]{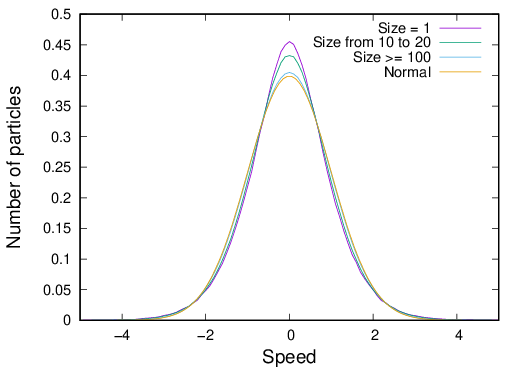}
\caption{$x$ component of the velocity distribution function at $t = 100$ for different sizes,  compared to the Maxwellian (normal) distribution. All velocity distributions are rescaled to $T = 1$. The typical size of clusters at $t=100$ is about $k=100$. }
\end{subfigure}
\hfill
\begin{subfigure}[t]{0.45\textwidth}
\centering
\includegraphics[width=\columnwidth]{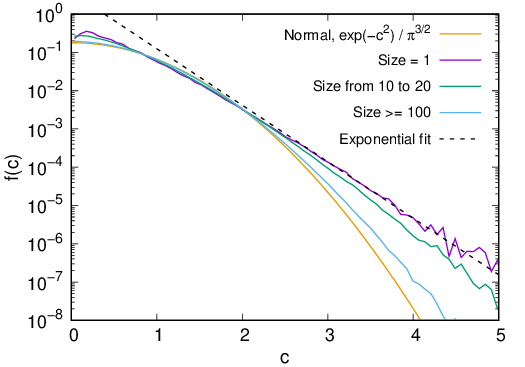}
\caption{Dimensionless spherically symmetric velocity distribution at $t = 100$ for various cluster sizes compared to Maxwell distribution and exponential tail.}
\end{subfigure}
\caption{Comparison between the Maxwell distribution and velocity distributions observed in Monte Carlo simulations of the aggregating system for the initial condition of $10^8$ monomers.}
\label{fig-sonine}
\end{figure}

Generally, the Maxwell distribution holds only for systems with bouncing collisions of particles, without energy loss. Inelastic collisions result in deviations from the Maxwell distribution, including the exponential high-velocity tails, e.g. \cite{BrilliantovPoeschelOUP}. However, aggregation is not just a special case of inelastic collisions. Intuitively, one can still expect that the Maxwell distribution may remain a good approximation for aggregating systems. Indeed, once the particles aggregated, the speed of the new cluster is the speed of the center of mass of the old pair. And if one takes some sample of random pairs of clusters, with Maxwell distribution, the centers of masses of the pairs will also obey the Maxwell distribution. Hence, the aggregation preserves the form of the Maxwell distribution, provided all clusters have the same temperature. The difference in temperatures may, however, affect the distribution and distort the Maxwellian; still, we expect that the distortion would be small.

\begin{figure}[!ht]
\includegraphics[width = \columnwidth]{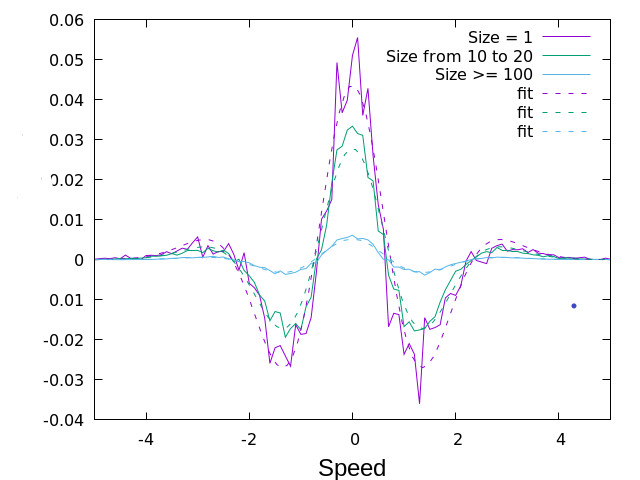}
\caption{The DSMC results for the absolute difference between the actual and Maxwellian VDF for $x$-component of  clusters velocity for clusters of different size at time $t=100$ (solid lines). The fitting of the difference with the second Sonine polynomial for $x-$ component (associated with $a_2$) is also shown (dashed lines). The initial number of monomers was $3\cdot 10^7$; the data is re-scaled for $T=1$.  Note that the deviations from the Maxwellian VDF can be well described by only one orthogonal polynomial. }
\label{fig-sonineDiff}
\end{figure}

To test our hypothesis of closeness between the real velocity distribution during aggregation and the Maxwell distribution, we perform Direct Simulation Monte Carlo \cite{dsmc} of the space homogeneous system, which obeys the equation \eqref{eq:boltz}:
\[
  \frac{d}{dt} f_k \left( \vec V_k \right) = \sum\limits_{i+j = k} I_{ij}^{{\rm agg},1} \left( \vec V_k \right) - \sum\limits_{j = 1}^{\infty} I_{kj}^{{\rm agg},2} \left( \vec V_k \right)
\]
with all collisions leading to aggregation, $\vec u_k = 0$ for all $k$, monodisperse initial conditions $n_1 = 1$ and $N_p = 3 \cdot 10^8$ initial particles with unit diameter and speeds from the Maxwell distribution with initial temperature $T_1 = 1$. Here we keep track of the velocity of each individual particle, exactly as in \cite{dsmc}.

Irreversible aggregation quickly leads to a multi-component system with the  so-called scaling distribution of particle sizes \cite{Leyvraz2003}. For aggregation described by the  Boltzmann equations, the average temperature scales as $T \sim t^{-1/3}$ and typical mass $s$ (scaling parameter) scales as $s = n^{-1} \sim t^{-1}$ \cite{equaltemporig}. So, at $t = 100$, the typical mass in the system is about 100. At this point monomers comprise less than $1\%$ of the system, and the difference from the Maxwell distribution for them is not significant, so they would not significantly affect the overall evolution of the system, which is mainly determined by particles of (large) typical size at this time. Recall that we are interested in the hydrodynamic stage   of a system evolution, when initial conditions are already forgotten. Hence, a relatively wide distribution of clusters' size emerges. A new cluster, say of  size $K$,  may be generally formed in many possible ways, from smaller clusters of different sizes and velocities. For  $K\gg 1$ we have a situation, somewhat similar to the  central limit theorem (although, rigorously speaking, not the same), which implies a normal (or Maxwellian) VDF for such clusters.

Fig.  \ref{fig-sonine} shows the velocity distribution, obtained by DSMC at the time $t=100$. The figure demonstrates that while the distribution deviates for small clusters from the Maxwellian at small velocities, it is very close to this  for large clusters. That is, the clusters of typical mass (about $100$ at $t=100$) or larger have the velocity distribution very close to the Maxwellian. This justifies the assumption of Maxwell distribution exploited in the main text in the derivation of Smoluchowski-Euler equations.

As we briefly mention in the main text and discuss in detail in the previous section of SM, the deviations from the Maxwellian distribution stem from the fields' gradients and inelasticity of collisions. They may be described within the 14-moment Grad's approach. The deviations due to inelasticity (aggregation in our case) are described by the fourth moment coefficient $a_2^{(k)}$. One can also use high-order moments in this expansion, that is, $a_3^{(k)}$, $a_4^{(k)},\ldots$; in most cases, however, it is sufficient to use only the first coefficient $a_2$ \cite{BrilliantovPoeschelOUP}, which is illustrated in Fig. \ref{fig-sonineDiff}. Furthermore, the deviations from the Maxwellian, Fig. \ref{fig-sonineDiff},  have positive and negative parts. These parts yield compensating contributions to the moments of the VDF -- the hydrodynamic fields, which additionally explains, why the  Maxwellian approximation for the VDF works so well.  (Note that in the quasi-one-dimensional case, as in the vertical fall model of the main text, we have one-dimensional analogues of Sonine polynomials associated with the vertical velocity $v_z$).

\begin{figure}[!ht]
\includegraphics[width=\columnwidth]{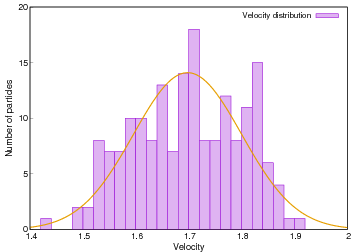}
\caption{VDF (bar charts) --   $z$ component of VDF for clusters of size $5$ for the case of vertical fall (dust sedimentation with aggregation model of the main text) at the last layer $z = -250 \, m$. There are in total about $5000$ clusters of all sizes at each of $100$ layers. Lines --  the corresponding Maxwell VDF.}
\label{fig:histoa}
\end{figure}

As another example, we consider VDF for the   case of vertical fall of the main manuscript for  clusters of size 5. These clusters appear only in pairwise collisions: $[2]+[3]\to [5]$ and $[1]+[4]\to [5]$. For the initial stage of the process (before the hydrodynamic stage develops) one may expect a VDF with two peaks, associated with the average velocities of corresponding pairs. This is indeed observed for the initial stage at the height $z=-50\, m$. However, already at the height $z = -250 \,m$, where the aggregation process is at the hydrodynamic stage,  a relatively smooth distribution is observed, see Fig. \ref{fig:histoa}. The smoothing effect at this height stems from the increasing velocity variance from collisions (i.e. clusters of size $2$, $3$ and $4$ enhance the velocity variance in preceding collisions) and the air friction. Hence, as the process develops  (clusters fall lower and lower), the distribution becomes smoother and closer to Maxwellian.

Moreover, as one can see from Fig. 1 of the main text, even noticeable deviations from the Maxwellian (at small heights, as for $z=-50\,m$) are not important. The VDF may be successfully replaced by the Maxwell distribution, leading to the same predictions for the hydrodynamic field. This may be explained as follows. The hydrodynamic equations depend on the first five moments of the VDF: density (zero moment), average velocity (three first moments) and velocity variance (i.e. temperature  -- the second moment), and the Maxwell distribution has already enough parameters to correctly capture and predict these moments. Certainly, one can consider more moments (e.g. 14-moment) and use more complicated VDF, but that would not affect the results significantly, as there are already almost no observable differences between the theory based on the Maxwellian VDF and DSMC results. Similar analysis has been performed for the case of spherically symmetric explosion, where we have not observed any double-peak distribution. This  may be explained by the lack of size-dependent air friction, which makes the  average velocities of clusters of different sizes different during the vertical fall.

\end{document}